\documentclass[]{aa}  

\usepackage{graphicx}
\usepackage[varg]{txfonts}
\usepackage{adjustbox}
\usepackage{parskip}
\usepackage[allcolors=blue,]{hyperref}

\makeatletter
\renewcommand*\aa@pageof{, page \thepage{} of \pageref*{LastPage}}
\makeatother

\begin{document}

\title{Signatures of anti-social mass-loss in the ordinary Type II SN 2024bch}
\subtitle{A non-interacting supernova with early high-ionisation features}

\newcommand{\ha}{H$\alpha$}
\newcommand{\hb}{H$\beta$}
\newcommand{\hg}{H$\gamma$}
\newcommand{\hd}{H$\delta$}
\newcommand{\kms}{\,km\,s$^{-1}$}
\newcommand{\msun}{\,M$_{\sun}$}
\newcommand{\ang}{\,\AA}
\newcommand{\ergs}{\,erg\,s$^{-1}$}
\newcommand{\sn}{SN~2024bch}

\author{
L.~Tartaglia\inst{1}\orcid{0000-0003-3433-1492} \and 
G.~Valerin\inst{2}\orcid{0000-0002-3334-4585} \and 
A.~Pastorello\inst{2}\orcid{0000-0002-7259-4624} \and 
A.~Reguitti\inst{3,2}\orcid{0000-0003-4254-2724} \and
S.~Benetti\inst{2}\orcid{0000-0002-3256-0016} \and 
L.~Tomasella\inst{2}\orcid{0000-0002-3697-2616} \and
P.~Ochner\inst{2,4}\orcid{0000-0001-5578-8614}
\and
E.~Brocato\inst{1,5}\orcid{0000-0001-7988-8177} \and
L.~Cond\`o\inst{6}\orcid{0009-0004-6220-6378} \and
F.~De~Luise\inst{1}\orcid{0000-0002-6570-8208} \and
F.~Onori\inst{1}\orcid{0000-0001-6286-1744} \and
I.~Salmaso\inst{7,2}\orcid{0000-0003-1450-0869}
}

\institute{
INAF -- Osservatorio Astronomico d'Abruzzo, via Mentore Maggini snc I-64100 Teramo, Italy \\ \email{leonardo.tartaglia@inaf.it} \and 
INAF -- Osservatorio Astronomico di Padova, vicolo dell'Osservatorio 5 I-35122 Padova, Italy \and 
INAF -- Osservatorio Astronomico di Brera, via E. Bianchi 46 I-23807, Merate, Italy \and
Dipartimento di Fisica e Astronomia, Universit\`a degli Studi di Padova, Via F. Marzolo 8, I-35131 Padova, Italy \and
INAF - Osservatorio Astronomioco di Roma (OAR), via Frascati 33, 00078 Monte Porzio Catone (RM), Italy \and
Dipartimento di Scienze Fisiche e Chimiche, Universit\`a degli Studi dell’Aquila, via Vetoio 42 I-67100 L’Aquila, Italy \and
INAF -- Osservatorio Astronomico di Capodimonte, Salita Moiariello 16, I-80131 Napoli, Italy
}

\date{Received October 26, 1985 01:21; accepted November 12, 1955 06:15}

\abstract{In this paper we analyse the spectro-photometric properties of the Type II supernova \sn, exploded in NGC~3206 at a distance of $19.9\,\rm{Mpc}$. Its early spectra are characterised by narrow high-ionisation emission lines, often interpreted as signatures of ongoing interaction between rapidly expanding ejecta and a confined dense circumstellar medium.
However, we provide a model for the bolometric light curve of the transient that does not require sources of energy different than radioactive decays and H recombination. Our model can reproduce the bolometric light curve of SN~2024bch adopting an ejected mass of $M_{bulk}\simeq5$\msun~surrounded by an extended envelope of only 0.2\msun~with an outer radius $R_{env}=7.0\times10^{13}\,\rm{cm}$. An accurate modelling focused on the radioactive part of the light curve, which accounts for incomplete $\gamma-$ray trapping, gives a $^{56}\rm{Ni}$ mass of 0.048\msun. We propose narrow lines to be powered by Bowen fluorescence induced by scattering of \ion{He}{II} Ly$\alpha$ photons, resulting in the emission of high-ionisation resonance lines. Simple light travel time calculations based on the maximum phase of the narrow emission lines place the inner radius of the H-rich, un-shocked shell at a radius $\simeq4.4\times10^{15}\,\rm{cm}$, compatible with an absence of ejecta-CSM interaction during the first weeks of evolution. Possible signatures of interaction appear only $\sim69\,\rm{days}$ after explosion, although the resulting conversion of kinetic energy into radiation does not seem to contribute significantly to the total luminosity of the transient.}

\titlerunning{Antisocial mass-loss in \sn}
\authorrunning{L.~Tartaglia et al.}

\maketitle

\section{Introduction} \label{sec:intro}
\begin{figure}
\begin{center}
\includegraphics[width=0.82\columnwidth]{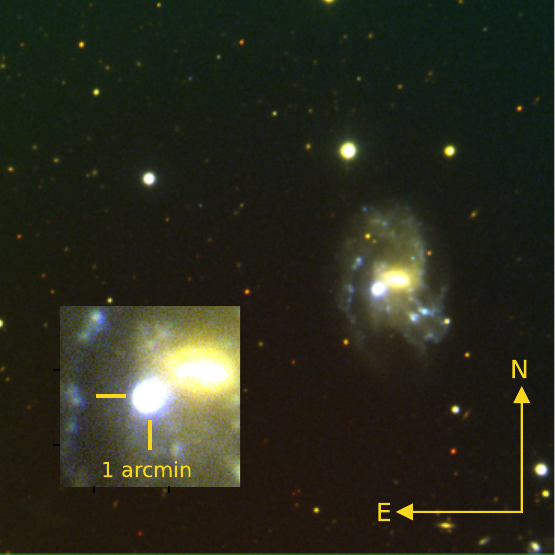}
\caption{Colour image of \sn~and its host galaxy NGC 3206, obtained combining $u-$, $g-$, $r-$ and $i-$band data obtained on 2024 February 5 with the $1.82\,\rm{m}$ Copernico telescope with AFOSC. The transient is the bright source in the middle of the inset. \label{fig:finder}}
\end{center}
\end{figure}
Core-collapse supernovae (CC SNe) are the endpoint of the evolution of massive stars \citep[$M\gtrsim8-9$\msun;][]{2003ApJ...591..288H}.
Hydrogen-rich CC SNe (labelled ``Type II" SNe) are the most common kind of explosion, with a further classification based on their photometric evolution, either showing a ``plateau" (Type IIP) or a ``linear" (Type IIL) decline after maximum light \citep{2019A&A...631A...8H}. 
Linearly evolving Type II SNe have been suggested to arise from stars partially depleted of their outer H layer, with relatively large radii \citep[$\sim10^3\,\rm{R_\sun}$;][]{1993A&A...273..106B} compared to the more compact red supergiant (RSG) progenitors associated to Type IIP SNe \citep[see, e.g.,][]{2009ARA&A..47...63S}. 

Narrow emission lines (full-width-at-half-maximum -- FWHM -- $10^2-10^3$\kms) are generally associated to SNe exploding within an extended circumstellar medium (CSM), where such features dominate the observed spectra throughout the entire evolution of the transients, as in the case of the H--rich Type IIn \citep[e.g.,][]{1990MNRAS.244..269S,1994ApJ...420..268C} and He--rich Type Ibn SNe (e.g., \citealt{2016MNRAS.456..853P}; see also \citealt{2017hsn..book..403S} and \citealt{2017hsn..book..843B} for more recent reviews).
The collision of the rapidly expanding SN ejecta ($\sim10^4$\kms) with the surrounding medium and the subsequent efficient conversion of kinetic energy into radiation may also contribute significantly to the energy output of these transients and produce slow-evolving transients \citep[see, e.g.,][]{2013A&A...555A..10T,2014ApJ...797..118F,2020A&A...635A..39T,2020A&A...638A..92T}.

Early observations \citep[a few hours to days after explosion; see, e.g.,][and references therein]{2021ApJ...907...52T} of core-collapse (CC) SNe may also show narrow, high-ionisation lines (e.g., \ion{He}{II}, \ion{N}{III} and, occasionally, \ion{C}{III/IV} and \ion{N}{IV/V}, \ion{O}{IV/V}; see \citealt{2014Natur.509..471G,2017NatPh..13..510Y,2023ApJ...952..119B}).
These features are believed to arise from shells expelled during the very last evolutionary stages of the progenitors and hence may reflect physical and chemical properties of their outer layers.
In this scenario, early narrow features would be the result of the photo-ionisation from the SN shock breakout \citep[the first electromagnetic signal from a SN; e.g.,][]{1992ApJ...393..742E,2011ApJS..193...20T}.
The presence of a surrounding shell may extend drastically the duration of the shock breakout signal, which, in Type II SNe, is expected to have a timescale of a few hundreds of seconds at most \citep[see, e.g.,][and references therein]{2015A&A...575L..10M}, depending on the radius of the progenitor star.  
An extension of a few days may have different explanations, including photon diffusion in a dense CSM \citep[e.g.,][]{2011MNRAS.414.1715B} and an aspherical explosion \citep[e.g.,][]{2010ApJ...717L.154S}.
Energetic photons produced in shocked regions may also help in increasing the lifespan of narrow lines, until the emitting shell is reached and swept away by the expanding SN ejecta.
This scenario, on the other hand, should also result in a drastic change in the profiles of narrow features, which is typically not observed.
Inflated envelopes have also been invoked to explain the relatively long duration of the shock breakout signal of the Type Ib SN~2008D \citep{2008Natur.453..469S,2009ApJ...702..226M}.
\citet{2015A&A...575L..10M} argued that such an extension may also affect the early spectroscopic evolution of SNe arising from H-rich progenitors. 

High-ionisation features were observed in early/very early spectra of Type II SNe, including SNe~1983K \citep{1985ApJ...289...52N,1990PASP..102..299P}, 1998S \citep{2000ApJ...536..239L,2015ApJ...806..213S}, 2006bp \citep{2007ApJ...666.1093Q}, PTF11iqb \citep{2015MNRAS.449.1876S}, LSQ13fn \citep{2016A&A...588A...1P}, 2013fs \citep{2017NatPh..13..510Y}, 2014G \citep{2016MNRAS.462..137T}, 2016bkv \citep{2018ApJ...861...63H}, 2017ahn \citep{2021ApJ...907...52T} 2020pni \citep{2022ApJ...926...20T} and 2020tlf \citep{2022ApJ...924...15J}.
More recent and well-studied transients showing similar early features include SN~2022jox \citep{2024ApJ...965...85A} and the nearby 2023ixf \citep[e.g.,][]{2023ApJ...954L..42J,2023ApJ...956...46S} and thanks to modern dedicated surveys, such as Distance Less Than 40 Mpc \citep[DLT40;][]{2018ApJ...853...62T} and Zwicky Transient Facility \citep[ZTF;][]{2019PASP..131g8001G,2019PASP..131a8002B,2019PASP..131a8003M}, the number of discoveries has increased dramatically in recent years.
\citet{2023ApJ...952..119B} claimed that the rate of SNe showing such features is expected to be relatively high (>30\% at 95\% confidence level).
\begin{table*}
\caption{Log of the spectroscopic observations of \sn.} \label{tab:speclog}
\centering
\resizebox{0.95\textwidth}{!}{
\begin{adjustbox}{tabular=cccccccc,center} \\
\hline\hline
Date & JD & Phase  & Telescope+Instrument & Grism/grating & Range & Resolution              & Exposure  \\
\hline
& & (days) & & & \AA & ($\lambda/\Delta\lambda$) & (s) \\
\hline
20240129 & 2460339.45 & $+1.40$ & Ekar182+AFOSC & Gr04 & $3600-8200$ & 310 & $1200$ \\
20240129 & 2460339.47 & $+1.42$ & Ekar182+AFOSC & VPH6+VPH7 & $3600-9200$ & $380+420$ & $900+900$ \\
20240130 & 2460340.48 & $+2.44$ & Ekar182+AFOSC & VPH6+VPH7 & $3600-9200$ & $380+420$ & $1800+1800$ \\
20240201 & 2460341.50 & $+3.45$ & Ekar182+AFOSC & VPH6+VPH7 & $3300-8800$ & $380+420$ & $2\times(1200+1200)$ \\
20240201 & 2460341.56 & $+3.51$ & Ekar182+AFOSC & VPH4 & $6300-7000$ & 3500 & $2\times1800$ \\
20240204 & 2460345.33 & $+7$ & T122+B\&C & 300tr & $3300-7400$ & 740 & $3\times1800$ \\
20240205 & 2460346.38 & $+8$ & Ekar182+AFOSC & VPH6+VPH7 & $3300-8800$ & $380+420$ & $1200+1200$ \\
20240213 & 2460354.46 & $+16$ & T122+B\&C & 600tr & $5000-7400$ & 1900 & $3\times1800$ \\
20240310 & 2460380.34 & $+42$ & Ekar182+AFOSC & VPH7 & $3600-7200$ & 420 & $900$ \\
20240317 & 2460386.51 & $+48$ & TNG+DOLoRes & LRB+LRR & $3600-9200$ & $580+710$ & $900+900$ \\
20240406 & 2460407.45 & $+69$ & Ekar182+AFOSC & VPH7 & $3800-7200$ & 250 & 900 \\
20240412 & 2460413.49 & $+75$ & Ekar182+AFOSC & VPHD1 (blue+red) & $3600-9200$ & $1200+730$ & 2700 \\
20240515 & 2460445.83 & $+107$ & TNG+DOLoRes & LRB & $3600-8100$ & 380 & 1800 \\
20240515 & 2460446.47 & $+108$ & TNG+DOLoRes & VHRV+VHRR & $4800-7800$ & $1530+2510$ & $2700+2700$ \\
20240616 & 2460476.40 & $+138$ & TNG+DOLoRes & LRB+LRR & $3600-9200$ & $580+710$ & $1200+1200$  \\ 
20240630 & 2460492.42 & $+154$ & TNG+DOLoRes & LRB & $3600-8100$ & 580 & 2700 \\
\hline
\end{adjustbox}}
\tablefoot{Ekar182: $1.82\,\rm{m}$ Copernico telescope at the Asiago observatory (Mount Ekar) with the ``Asiago Faint Object Spectrograph and Camera" (AFOSC); T122: $1.22\,\rm{m}$ Galileo telescope at the Osservatorio Astrofisico di Asiago; TNG: $3.58\,\rm{m}$ Telescopio Nazionale Galileo at La Palma (Canary Islands, Spain) with the ``Device Optimised for the Low Resolution'' (DOLoRes) camera. Phases refer to the estimated epoch of the SN explosion.}
\end{table*}

Here we report results of our analysis on UV/Optical data of \sn, obtained soon after discovery and covering up to the early nebular phases (corresponding to the first seasonal gap).
Analysis on later data will be performed and presented in a forthcoming paper, once \sn~will be observable again using ground-based facilities.
The transient was discovered on 2024 January 29.27~UT \citep{2024TNSTR.281....1W} in the nearby SB(s)cd NGC~3206 \citep{1991rc3..book.....D}, at $\alpha=$10:21:49.740, $\delta=+$56:55:40.51 [J2000] (see Fig.~\ref{fig:finder}), which, at a distance of $19.9\pm4.1\,\rm{Mpc}$ \citep{2016AJ....152...50T} and adopting an inclination of 61\farcs1 \citep[as reported in the HyperLeda database\footnote{\url{http://leda.univ-lyon1.fr}};][]{2014A&A...570A..13M} corresponds to a de-projected distance of $3.9\,\rm{kpc}$ from the host centre.
The SN was confirmed by ongoing transient surveys\footnote{\url{https://www.wis-tns.org/object/2024bch}} such as the Gravitational-wave Optical Transient Observer \citep[GOTO;][]{2022MNRAS.511.2405S}, the Asteroid impact early warning system \citep[ATLAS;][]{2020PASP..132h5002S}, ZTF and the Panoramic Survey Telescope \& Rapid Response System \citep[Pan-STARRS;][]{2016arXiv161205560C} within the Young Supernova Experiment \citep[YSE;][]{2021ApJ...908..143J} and assigned the internal names GOTO24hm, ATLAS24bmx, ZTF24aaghpeh and PS24aap, respectively. 
Additional photometric points were later provided by the Mobile Astronomical System of Telescope-Robots \citep[MASTER;][]{2004AN....325..580L} and by the ESA Gaia Photometric Science Alerts Team\footnote{\url{http://gsaweb.ast.cam.ac.uk/alerts)}}, who further labelled the transient MASTER~OT~J102149.74+565540.4 and Gaia24bdk. 
Low-resolution spectroscopy was used to classify \sn~as a young SN with high-ionisation features \citep{2024TNSCR.284....1B,2024TNSCR.465....1M}. 
We then started our UV/optical follow-up campaign detailed in Sect~\ref{sec:observations}, revealing a spectro-photometric evolution consistent with that of a Type II SN, with narrow features disappearing a few days after discovery and little or no contribution from a source of energy in addition to the ``canonical" H recombination and radioactive decays, as we discuss in Sect.~\ref{sec:analysis}.
A summary of our main results is reported in Sect.~\ref{sec:conclusions}.

\section{Observations and data reduction} \label{sec:observations}
The follow-up campaign of \sn~started soon after discovery \citep[corresponding to 2024 January 29.29~UT;][]{2024TNSTR.281....1W}, with early photometry obtained on 2024 January 29.92~UT and low-resolution spectroscopy on 2024 January 29.95~UT (see Table~\ref{tab:speclog}).
Optical observations were mostly carried out with facilities of the INAF - Osservatorio Astronomico di Padova (i.e., the $1.82\,\rm{m}$ Copernico with the Asiago Faint Object Spectrograph and Camera -- AFOSC -- and the $69/92\,\rm{cm}$ Schmidt with a G4-16000LC Moravian camera, both located in Asiago).
Additional $ugriz$ photometry were provided by the Wide-field Optical Telescope (WOT), a $67/91\,\rm{cm}$ Schmidt telescope equipped with an Apogee Aspen CG16M camera located at the Campo Imperatore observatory in l'Aquila (Italy).
These data were reduced using the dedicated pipeline SuperNOva PhotometrY ({\sc SNOoPY}\footnote{\url{https://sngroup.oapd.inaf.it/snoopy.html}}) fitting the point spread function (PSF) computed on selected non-saturated stars in the field and zero-point calibration using stars from the Sloan Digital Sky Survey (SDSS) catalogue \citep[Data Release 17 -- DDR17;][]{2022ApJS..259...35A}. Further details on the reduction steps within {\sc SNOoPY} are available in \citet{2016PhDT.......149T}.

Near ultra-violet (NUV) observations were carried out using the $0.3\,\rm{m}$ Ultra-Violet Optical Telescope (UVOT) on board the Swift Gamma Ray Burst Explorer \citep{2004AAS...20511601G}.
These were reduced using {\sc HEASoft v. 6.33} \citep{2014ascl.soft08004N} on pre-processed images retrieved from the Swift archive\footnote{\url{https://swift.gsfc.nasa.gov/archive/}}, following the prescription of \citet{2009AJ....137.4517B}.
The field of \sn~was also monitored by ZTF and we hence used its forced-photometry service \citep{2019PASP..131a8003M} to collect extra $g-$ and $r-$band epochs, spanning a period of $\sim4\,\rm{years}$ prior discovery.
UV and $UBV$ photometry were both calibrated in the Vega photometric system, while $ugriz$ magnitudes were referred to the AB system.

Optical spectroscopy was reduced using the {\sc foscgui} pipeline\footnote{\url{https://sngroup.oapd.inaf.it/foscgui.html}} designed for the reduction of AFOSC data and performing standard {\sc iraf} \citep{1986SPIE..627..733T,1993ASPC...52..173T} reduction steps through {\sc pyraf} \citep{2012ascl.soft07011S}.

\begin{figure*}
\begin{center}
\includegraphics[width=\linewidth]{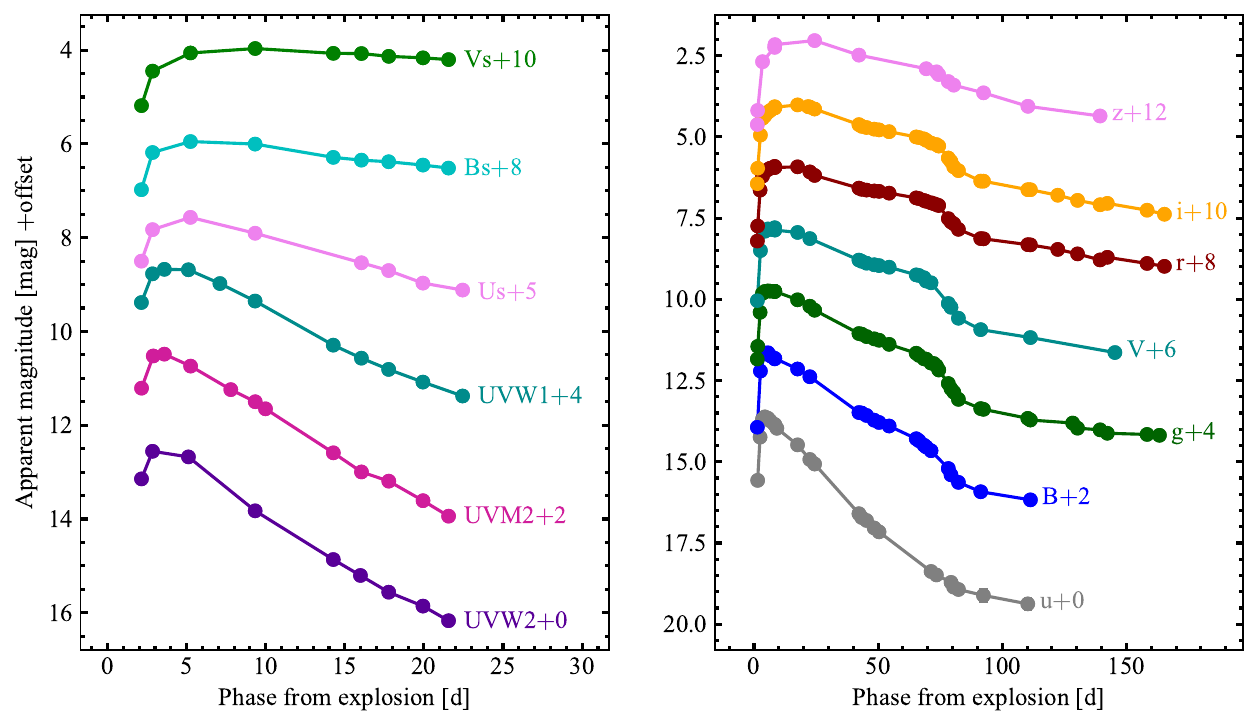} 
\caption{UVOT ($w2,\,m2\,w1\,u\,b\,v$; {\bf left}) and ({\bf right}) optical $ugBVriz$ light curves of \sn. Magnitudes were not corrected for extinction. UV and $BV$ magnitudes were calibrated in the Vega and $griz$ magnitudes in the AB photometric systems. \label{fig:lightCurves}}
\end{center}
\end{figure*}

\section{Analysis and modelling} \label{sec:analysis}
\subsection{Photometric evolution} \label{sec:photometry}
The photometric monitoring of \sn~started soon after discovery \citep[$\simeq0.4\,\rm{days}$ after the report by][see Section~\ref{sec:intro}]{2024TNSTR.281....1W}.  
The GOTO collaboration later provided an earlier discovery on 2024 January 29.05~UT ($\sim0.5\,\rm{days}$ earlier), with a non-detection on 2024 January 28.04~UT, corresponding to $\rm{JD}=2460337.5$. 
Taking the midpoint between the GOTO reported epochs, we then assume the explosion occurred on $\rm{JD}=2460338.0\pm0.5$ and refer phases to this epoch. Forced photometry on archival ZTF images \citep{2019PASP..131a8003M} did not show evidences of variability down to limiting magnitudes $g>20.5$, $r>20.6\,\rm{mag}$ $\sim4\,\rm{years}$ prior discovery, ruling out pre-SN outbursts with an absolute peak magnitude brighter than $\simeq-11\,\rm{mag}$.

Details on the instruments used and data reduction can be found in Sect.~\ref{sec:observations} while light curves are shown in Figure~\ref{fig:lightCurves}.
The early evolution is fast, with rise-times spanning from $\simeq4.4$ to $\simeq16.3\,\rm{days}$, in $u-$ and $z-$band, respectively, estimated using 5th order polynomials to the early light curve and adopting the explosion epoch reported above. 
Within the first $\sim1.6\,\rm{days}$ after explosion, optical magnitudes rise with an average rate of $3\,\rm{mag}\,\rm{day^{-1}}$, confirming that \sn~was discovered very soon after explosion. 

Following \citet{2014ApJ...786...67A} \citep[see also][]{2014ApJ...786L..15G}, we infer $V-$band photometric quantities to be compared with those available in the literature (see Table~\ref{tab:lcparams}). 
These include the end of the initial steeper decline of the plateau phase $t_{tran}$, the mid-point of the plateau to linear decline transition $t_{PT}$, the end of the plateau phase $t_{end}$ and the corresponding absolute magnitude $M_{end}$ \citep[see, e.g., Fig.~1 in][]{2014ApJ...786...67A}.
We could identify the two distinct phases of the decline after maximum, consisting in a fast steeper decline, followed by the slower plateau phase with slopes $s_1$ and $s_2$, respectively, while in agreement with \citet{2014ApJ...786...67A}, we refer to the slope of the radioactive tail using the $s_3$ parameter.
$t_{tran}$, $t_{end}$ and $t_{TP}$ were computed using the mid-points between last and first epochs of adjacent phases and their errors were estimated adding uncertainties on the explosion epoch and those due to gaps in the light curves (i.e., semi-amplitudes among adjacent phases) in quadrature. 
Adopting the most recent redshift-independent distance for NGC~3206 \citep[$19.9\pm4.1\,\rm{Mpc}$, corresponding to a distance modulus $\mu=31.49\pm0.45\,\rm{mag}$;][]{2016AJ....152...50T}, we infer a $V-$band peak absolute magnitude $-17.71\pm0.03\,\rm{mag}$ (where the uncertainty on the distance modulus of $0.45\,\rm{mag}$ was not included), with the maximum light occurring at $+7.7\,\rm{days}$. After maximum, at $t<t_{tran}=+44.4\pm2.0\,\rm{days}$ we infer $s_1=3.41\pm0.18\,\rm{mag}/100\,\rm{days}$ performing $10^4$ Monte Carlo simulations randomly shifting the data points within their errors. 
Mean values and standard deviations of the resulting distributions were taken for $s_1$, $s_2$ and $s_3$ and their uncertainties. 
At later phases, the light curve evolves slower, with a slope $s_2=1.83\pm0.18\,\rm{mag}\,\rm{/100\,\rm{days}}$, up to $t_{end}=67.5\pm1.1\,\rm{days}$, resulting in a length of the optically thick phase (``plateau duration") of $Pd=77.2\pm4.6\,\rm{days}$.
Both $s_1$ and $s_2$ are higher than mean values inferred by \citet{2014ApJ...786L..15G} for their sample of 52 Type II supernovae.
The derived $s_2$ value is also relatively large if compared to samples of well-studied SNe II \citep[see][]{2012ApJ...756L..30A,2015ApJ...799..208S,2016MNRAS.459.3939V}, in particular to those of \citet{2014MNRAS.442..844F} and \citet{2014MNRAS.445..554F}, who set the maximum slope for the plateau phase of Type IIP SNe to $\simeq0.25\,\rm{mag}/100\,\rm{days}$, suggesting \sn~belongs to the linearly declining Type II SNe class.
At $t\geq+67.5\,\rm{days}$ the light curve rapidly drops, with a decline of $\simeq1.7\,\rm{mag}$ within the following $\simeq25\,\rm{days}$. 
At $t\gtrsim91\,\rm{days}$, the $V-$band light curve settles on the radioactive tail, with a slope $s_3\simeq1.3\,\rm{mag}/100\,\rm{days}$.

\begin{figure}
\begin{center}
\includegraphics[width=\linewidth]{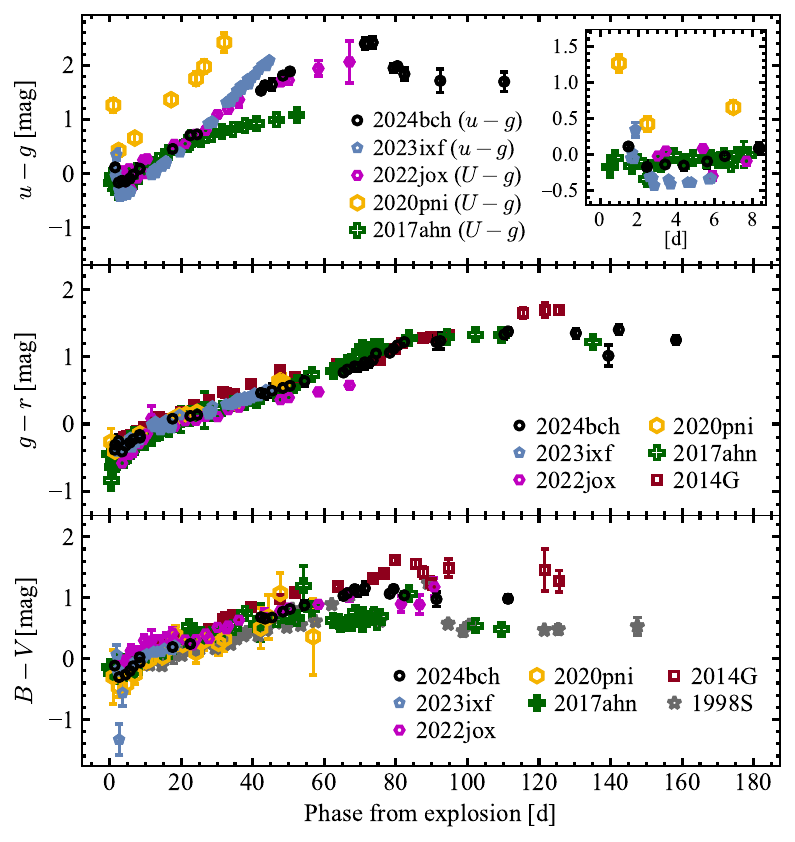}
\caption{Colour evolution of \sn~compared to the ones of selected transients showing high-ionisation features in their early spectra. Objects were selected among those with a similar evolution of the main features (including the duration of the narrow features and SN type) with data available in the literature. In the upper panel, the evolution of the $U-g$ colours for SNe~2017ahn, 2020pni and 2022jox was included, since $u-$band photometry for these objects was not available. $U-$band magnitudes for these objects were converted in the AB photometric system adopting the Vega - AB Magnitude conversions reported in \citet{2007AJ....133..734B}. In the same panel, the inset includes a zoom-in of the $u/U-g$ early evolution, showing a rapid decrease in the colours within the first $\sim3\,\rm{days}$ after explosion for all the selected objects (see the main text). Phases refer to the estimated explosion epochs reported in the literature. Light curves of SN~2023ixf were collected from \citet{2024Natur.627..754L} and \citet{2024Natur.627..759Z}. \label{fig:comparison}}
\end{center}
\end{figure}

In Fig.~\ref{fig:comparison}, we compare the photometric evolution of \sn~to those of a sub-sample of Type II SNe showing high-ionisation lines in their early spectra.
Among these, SNe~1998S, 2014G, 2017ahn, 2020pni, 2022jox and 2023ixf are the ones showing similar early features over a comparable period after estimated explosion epochs.
While the different temperature evolution is highlighted by the $u-g$ colours, we note a remarkable similarity among the $g-r$ and $B-V$ light curves.
In addition, the early $u-g$ evolution shows a rapid evolution from relatively red to blue colours within the first $\sim3\,\rm{days}$ after explosion in all selected objects (see the inset in Fig.~\ref{fig:comparison}).

\subsection{Modelling of the bolometric light curve} \label{sec:modelling}
The bolometric light curve was inferred fitting a blackbody to the spectral energy distribution (SED) computed at each epoch, providing estimates of the photospheric radius, temperature and corresponding luminosity.
At early epochs with available UV data ($t<+25\,\rm{days}$), we interpolated light curves at the epochs of the UV photometry, while at later times the interpolation was performed with respect to the $r-$band light curve (the best sampled one).
To avoid extrapolation, UV fluxes beyond $+25\,\rm{days}$ were exluded; however, UV flux estimated from the blackbod fit were incorporated into luminosity calculations via the Stefan-Boltzmann law.
The SEDs were inferred adopting zero points and $\lambda_{eff}$ reported in the Spanish Virtual Observatory \citep[SVO\footnote{\url{http://svo2.cab.inta-csic.es/theory/fps/}};][]{2012ivoa.rept.1015R,2020sea..confE.182R} database, referring to values in the AB or Vega systems, according to the calibration adopted for each filter (see Section~\ref{sec:observations}).
\begin{table}
\caption{$V-$band light curves parameters of \sn.}
\label{tab:lcparams}
\centering
\begin{tabular}{cc}
\hline\hline
Parameter & Value (err) \\
\hline
$t_0$ & 60337.5(0.5) \\
$M_V$ & $-17.71(0.03)\,\rm{mag}$ \\
$M_{end}$ & $-16.26(0.03)\,\rm{mag}$ \\
$M_{tail}$ & $-14.60(0.09)\,\rm{mag}$ \\
$t_{\rm{M}_V}$ & 60345.2(0.5) \\
$s_1$ & $3.41(0.18)\,\rm{mag/}100\,\rm{days}$ \\
$s_2$ & $1.83(0.18)\,\rm{mag/}100\,\rm{days}$ \\
$s_3$ & $1.31(0.23)\,\rm{mag/}100\,\rm{days}$ \\
$t_{tran}$ & $44.4(2.0)\,\rm{days}$ \\
$t_{end}$ & $67.46(1.14)\,\rm{days}$ \\
$Pd$ & $77.2(4.6)\,\rm{days}$ \\
\hline
\end{tabular}
\tablefoot{Reported epochs are modified Julian dates: MJDs.}
\end{table}
The resulting photometry was corrected for a Galactic reddening $E(B-V)=0.013\,\rm{mag}$ \citep{2011ApJ...737..103S}, considering a negligible contribution of the local environment to the total reddening, as we did not detect \ion{Na}{ID} absorption features at the host redshift in the SN spectra.
The resulting evolution of temperature, radius and bolometric luminosity assuming a distance of $19.9\pm4.1\,\rm{Mpc}$, is reported in Table~\ref{table:bolom}.
\begin{table}
\caption{Parameters of the blackbody fit to the SEDs of \sn.} \label{table:bolom}
\centering
\begin{tabular}{cccc}
\hline \hline
Phase  & $\rm{T_{ph}}$ (err) & $\rm{R_{ph}}$ (err) & $L_{bol}$ (err)  \\
(days) &  (K) & ($10^{14}\,\rm{cm}$) & ($10^{42}$\ergs) \\
         \hline
$+2.2$	 & 16289(75) & 2.506(0.026) & 3.15(0.10) \\
$+2.9$	 & 15163(88) & 3.658(0.041) & 5.04(0.16) \\
$+3.6$	 & 13663(73) & 4.59(0.10)   & 5.24(0.17) \\
$+5.3$	 & 11993(47) & 5.92(0.10)   & 5.17(0.13) \\
$+7.8$	 & 10338(40) & 7.49(0.10)   & 4.56(0.12) \\
$+9.3$	 & 9601(31)  & 8.24(0.10)   & 4.11(0.10) \\
$+10.0$	 & 9308(31)  & 8.59(0.10)   & 3.95(0.10) \\
$+14.3$	 & 7789(23)  & 10.66(0.11)  & 2.98(0.12) \\
$+16.1$	 & 7299(21)  & 11.62(0.12)  & 2.73(0.13) \\
$+17.8$	 & 6941(19)  & 12.37(0.13)  & 2.53(0.15) \\
\hline
\end{tabular}
\tablefoot{Evolution of the photospheric temperature and radius and the corresponding bolometric luminosity of \sn. Errors are the $1\sigma$ uncertainties derived by {\sc curve\_fit} (see the main text). Phases refer to the estimated epoch of the SN explosion. The Table is published in its entirety in the machine-readable format. A portion is shown here for guidance regarding its form and content.}
\end{table}

\begin{figure}
\begin{center}
\includegraphics[width=\columnwidth]{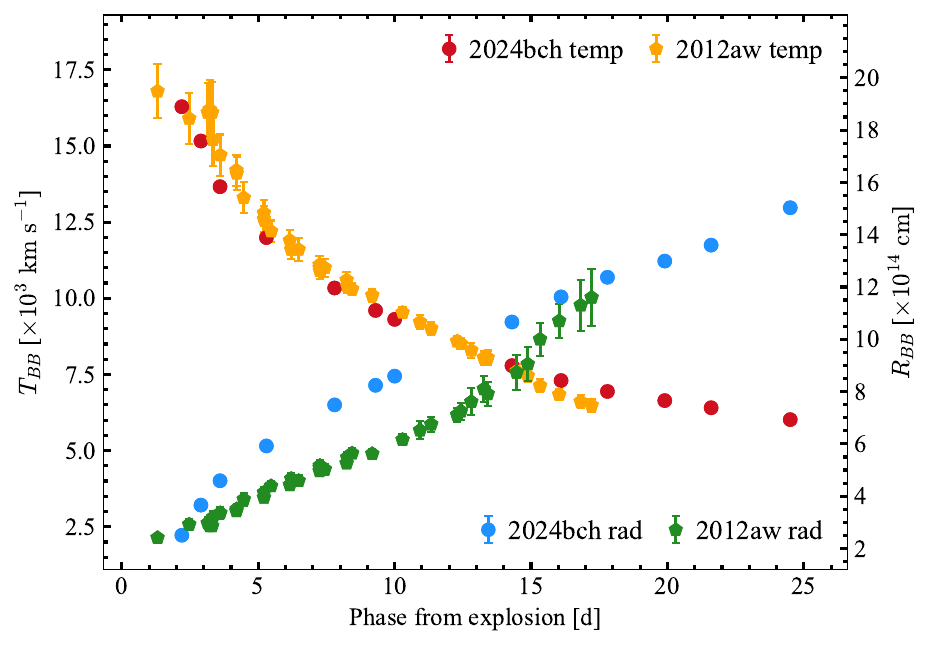}
\caption{Evolution of the radius and temperature inferred from the blacbody fit to the early SED of \sn~compared to the ones of SN~2012aw \citep[][]{2014ApJ...787..139D} computed using the same approach. \label{fig:tempradev}}
\end{center}
\end{figure}

The evolution of both the photospheric temperature and radius already suggests the presence of a relatively low-mass ejected shell which rapidly cools down, and little or no contribution from alternative sources of energy, such as efficient conversion of kinetic energy into radiation through shocks (see Fig.~\ref{fig:tempradev}).

\begin{figure}
\begin{center}
\includegraphics[width=\columnwidth]{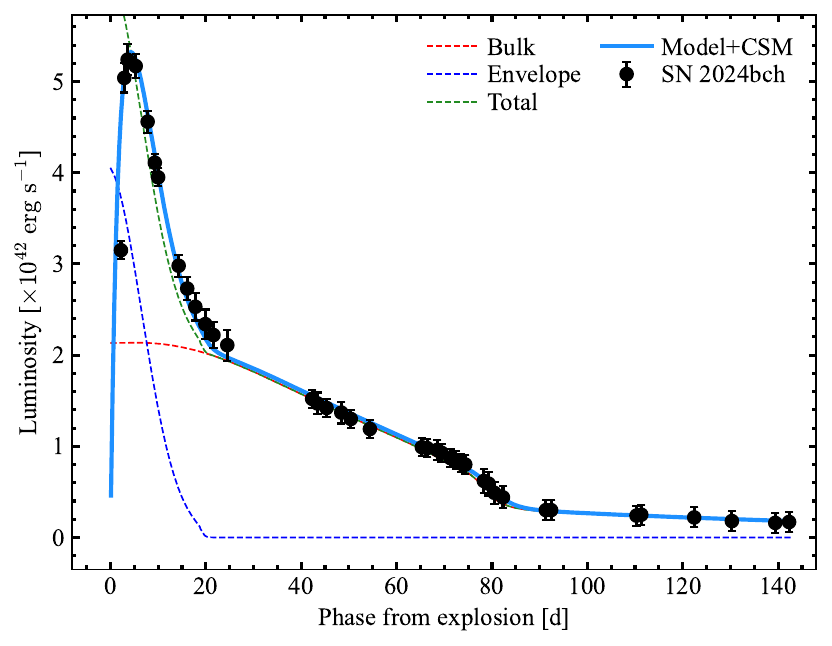}
\end{center}
\caption{Evolution of the bolometric luminosity of \sn~along with the model used to reproduce the observed data. Dashed lines represent the luminosity contribution from both the envelope and the bulk of the ejecta. The green solid line is the sum of the two components, while the light blue solid line shows the contribution of the extra diffusion time introduced by the CSM. \label{fig:ModelBolom}}
\end{figure}

\begin{table*}
\caption{Parameters used in the models displayed in Fig.~\ref{fig:ModelBolom}.}
\centering
\begin{tabular}{cccccccccc}
 \hline 
 \hline
  \textsuperscript{56}Ni & M(bulk) & E$_{k}$(bulk) & E$_{th}$(bulk) & R(bulk) & M(env) & E$_{k}$(env) & E$_{th}$(env) & R(env) & t$_{CSM}$ \\ 
  (M$_{\odot}$) & (M$_{\odot}$) & (erg) & (erg) & (cm) & (M$_{\odot}$) & (erg) & (erg) & (cm) & (days) \\ 
 \hline
  4.3x10$^{-2}$ & 5.0 & 1.65x10$^{51}$ & 5.0x10$^{50}$ & 3.5x10$^{13}$ & 0.2 & 7x10$^{49}$ & 1.9x10$^{49}$ & 7.0x10$^{13}$ & 1.4 \\
 \hline
\end{tabular}
\label{tab:ModelsParam}
\end{table*}
To explore this scenario, we tried to reproduce the bolometric light curve of \sn~using the model provided by \cite{2016A&A...589A..53N}. 
In this model, the light curve results from the contribution of two distinct regions: a massive ``bulk" of ejecta surrounded by a less massive envelope \citep[see also][]{2023A&A...673A.127S}. 
The model assumes homologous expansion of the ejecta, spherical symmetry, and constant opacity \citep[$0.34,\rm{cm^2},\rm{g^{-1}}$ for a H--rich medium with solar composition;][]{1986rpa..book.....R,2019arXiv191200844L}. 
Radiation transport is treated using the diffusion approximation, and the bolometric luminosity is only powered by radiation generated from $^{56}\rm{Ni}$ radioactive decay, energy released during gas expansion and cooling and H recombination \citep[see][]{1989ApJ...340..396A,2014A&A...571A..77N}. 
\citet{2016A&A...589A..53N} further assume a density structure for the supernova ejecta consisting of an inner region with constant density up to a dimensionless radius $x_0$, and an outer region characterised by either an exponential or power-law density profile. 
Furthermore, they report that a constant density profile for the outer H--rich envelope is particularly effective in fitting the plateau phase of many Type II SNe, consistent with the prescription of \citet{1989ApJ...340..396A}. 
For our analysis, we employed both a flat (constant) density profile and a power-law profile ($\rho \propto r^{-2}$) for the outer region \citep[see, e.g.,][]{2011MNRAS.415..199M}, noting that this choice minimally affects the overall fit, only influencing the derived envelope mass.
Following the formalism reported in \citet{2012ApJ...746..121C} (see their Eq.~4), we also introduced an additional diffusion time term ($t_{CSM}$), simulating the presence of an optically thick CSM along the line of sight. 
Although observational evidence suggests the need of a ``detached" CSM (e.g., early high-ionisation features and the observed rise in the bolometric light curve), we remark this component does not add a further powering mechanism \citep[see also][]{Valerin2024}: no extra energy is provided by ejecta-CSM interaction.
We found that the early, sharp peak of \sn~is compatible with the energy radiated by a 0.2\msun~cooling envelope characterised by a low diffusion time. The longer-lasting plateau phase is well constrained adopting 5\msun~of ejecta containing $4.3\times10^{-2}$\msun~of $^{56}\rm{Ni}$. 
Although this value does not take into account incomplete $\gamma-$ray trapping, it is comparable to the one discussed later in this Section, resulting from a more accurate analysis of the late light curve.
This is also similar to the mean value of $0.037\pm0.005$\msun~found by \citet{2021MNRAS.505.1742R} for their sample of 109 Type II SNe.
The velocities adopted for both the bulk and envelope are $\simeq7500$\kms, comparable to the ejecta velocities derived from the spectral analysis (see Section~\ref{sec:spectroscopy}). 
This is of paramount importance, given the partial degeneracy between the model parameters (e.g., a larger $v_{ej}$ would give a larger $M_{ej}$).
Velocity measurements from the minima of the P-Cygni features allow to us constrain the characteristic velocity of the gas, therefore allowing for a more reliable estimate of the ejected mass.
In Fig.~\ref{fig:ModelBolom}, the resulting model is compared to the observed bolometric luminosity of \sn, while the parameters used are reported in Table \ref{tab:ModelsParam}.
The best model was obtained by generating a grid of over 50 models and selecting the one with the lowest root mean square error with respect to the observed data.

\begin{figure}
\begin{center}
\includegraphics[width=\columnwidth]{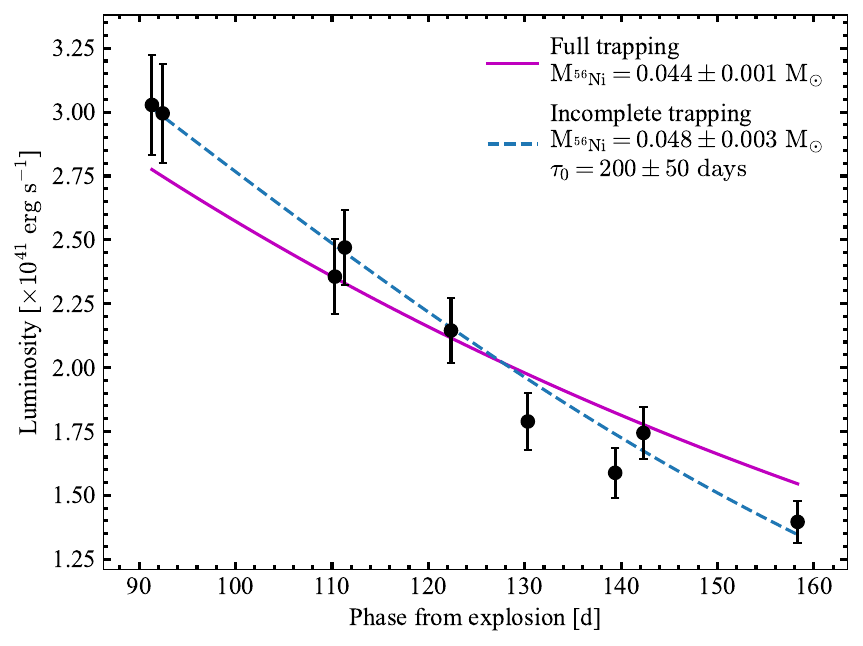}
\end{center}
\caption{Evolution of the late bolometric luminosity of \sn~with respect to the $r-$band light curve (see the main text), including the fit on the radioactive tail. The fit assuming full $\gamma-$ray trapping is also reported. \label{fig:lateBoloFit}}
\end{figure}

At $t\gtrsim+85\,\rm{days}$ the bolometric light curve settles on the radioactive tail, where the luminosity output is dominated by deposition and re-emission of the $\gamma-$rays produced in the reaction chain $^{56}\rm{Ni}\rightarrow{^{56}}\rm{Co}\rightarrow{^{56}\rm{Fe}}$.
With an e-folding time of $8.8\,\rm{days}$, the $^{56}\rm{Ni}$ to $^{56}\rm{Co}$ decay releases energy during the early phase of the light curve. Its luminosity contribution is delayed by the diffusion time within the bulk of the ejecta and mixed with the emission of the surrounding cooling gas.
$^{56}\rm{Co}$, instead, has a much longer lifetime \citep[$111.4\,\rm{days}$;][]{1994ApJS...92..527N,1999NDS....86..315J}, and the last part of the radioactive chain is hence expected to shape the late light curves of CC SNe, after the end of the recombination-dominated phase.
The predicted luminosity output during the $^{56}\rm{Co}-$decay phase is given by:
\begin{equation}
L=9.92\times10^{41}\frac{M_{\rm{^{56}Ni}}}{0.07\,\rm{M_{\odot}}}\left(e^{-t^{}/111.4\,\rm{d}}-e^{-t^{}/8.8\,\rm{d}}\right)\,\rm{erg}\,\rm{s^{-1}}.
\label{eq:lumtail}
\end{equation}
Equation~\ref{eq:lumtail} assumes full trapping of the $\gamma-$rays produced during the decays and instantaneous re-emission of the deposited energy \citep[][]{2012A&A...546A..28J}.
The late evolution of \sn, on the other hand, slightly deviates from that expected by Eq.~\ref{eq:lumtail} and seems to evolve faster (see Fig.~\ref{fig:lateBoloFit}).
A similar behaviour was observed in the stripped-envelope (SE) SNe~1983N, 1983V and 1993J \citep{1996ApJ...459..547C,1997ApJ...483..675C,1994AJ....107.1022R} and, more recently, in the linearly declining Type II SNe 2014G and 2017ahn \citep[][both showing a spectro-photometric evolution similar to \sn]{2016MNRAS.462..137T,2021ApJ...907...52T} and attributed to incomplete trapping of $\gamma-$rays produced in radioactive decays.
Following \citet{1997ApJ...491..375C}, we then included the additional term:
\begin{equation}
L(t)=L_0\,\left(1-e^{-\tau_0^{2}/t^2}\right),
\label{eq:gammaleak}
\end{equation}
where $L_0$ is given by Eq.~\ref{eq:lumtail} and $\tau_0$ is a full-trapping characteristic timescale which can be expressed as:
\begin{equation}
\tau_0=\left(C\kappa_\gamma\frac{M^2_{ej}}{E_k}\right)^{1^{}/2},
\label{eq:trapscale}
\end{equation}
with $M_{ej}$ and $E_k$ mass and kinetic energy of the SN ejecta, respectively, $\kappa_\gamma$ $\gamma-$ray opacity and $C$ a constant which can be expressed analytically by $C=(\delta-3)^2\left[8\pi\left(\delta-1\right)\left(\delta-5\right)\right]^{-1}$ for a radioactive medium with a density profile $\rho(r,t)\propto r^\delta(t)$.
Fitting Eq.~\ref{eq:gammaleak} to the late light curve of \sn, we inferred a $^{56}\rm{Ni}$ mass of $0.048\pm0.003$\msun~and $\tau_0=200\pm50\,\rm{days}$, where values and uncertainties were obtained performing $10^4$ Monte Carlo simulations.
Assuming a typical $E_k$ of $10^{51}\,\rm{erg}$, $\kappa_{\gamma}=0.06\,\rm{cm^2}\,\rm{g^{-1}}$ \citep{2015ApJ...814...63M} and $\delta=0$, including $\tau_0\simeq200\,\rm{days}$ in Eq.~\ref{eq:trapscale} then gives a $M_{ej}\simeq4.5$\msun.
Both $M_{^{56}\rm{Ni}}$ and $M_{ej}$ are in agreement with the masses predicted by our simple model described above (0.043 and 5.2\msun; see Table~\ref{tab:ModelsParam}).
The derived $^{56}\rm{Ni}$ mass, in particular, is very similar to the Ni mass found for the Type II SN~2022jox \citep[0.04\msun;][]{2024ApJ...965...85A} and consistent with mean values inferred by statistical studies on CC SNe (see the discussion above).

\subsection{Spectroscopic evolution} \label{sec:spectroscopy}
\begin{figure*}
\begin{center}
\includegraphics[width=0.85\linewidth]{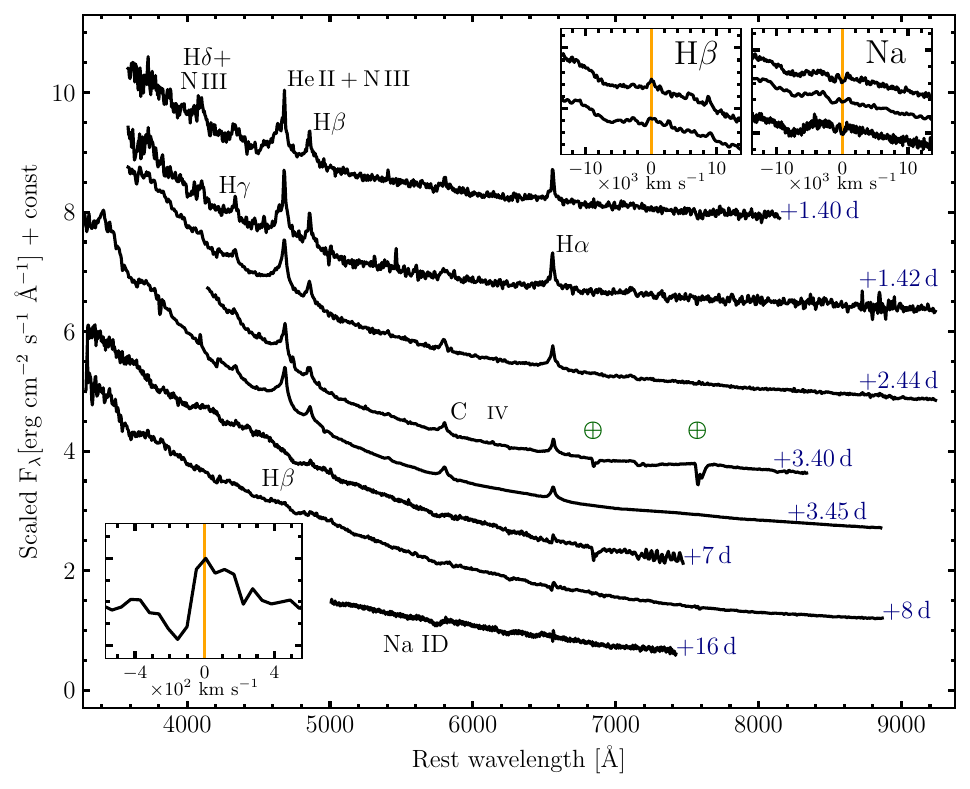}
\end{center}
\caption{Early spectroscopic evolution of \sn~(up to $+16\,\rm{days}$). Spectra were not corrected for Galactic extinction along the line of sight. $\oplus$ symbols mark the position of the main telluric features. Zoom-in panels show the \hb~and \ion{He}{I}/\ion{Na}{ID} regions (at $+7$ and $+8\,\rm{days}$ for \hb, and $+7$, $+8$ and $+16\,\rm{days}$ for \ion{Na}{ID}) and \ha, with expansion velocities referred to the rest-frame wavelengths marked with yellow vertical lines. Rest-frame phases refer to the estimated explosion epoch. \label{fig:earlyIDSpec}}
\end{figure*}

Early spectra ($t\gtrsim+1.4\,\rm{days}$; see Fig.~\ref{fig:earlyIDSpec}) show narrow emission lines over a blue continuum ($T\gtrsim1.8\times10^4\,\rm{K}$), similar to those observed in H-rich interacting SNe \citep[SNe IIn;][]{1990MNRAS.244..269S,2017hsn..book..403S}, with broad wings due to electron scattering occurring in a dense, relatively slow-moving CSM \citep[see][]{2001MNRAS.326.1448C,2018MNRAS.475.1261H}. 
The most prominent features are H Balmer lines (\ha, \hb, \hg~and \hd), \ion{He}{II} $\lambda4686$, \ion{N}{III} $\lambda4641$ as well as the \ion{C}{IV} doublet $\lambda\lambda5801$, 5812.
The presence of \ion{N}{III} $\lambda4641$ also suggests $\lambda4103$ likely contaminate the flux of \hg.
At $+2.4\,\rm{days}$ we also identify a faint feature corresponding to \ion{He}{II} $\lambda5412$. This is the only transition of the \ion{He}{II} Pickering-Fowler series \citep{2006agna.book.....O} not forming a blend with H Balmer lines, suggesting that \ion{He}{II} $\lambda6560$, $\lambda4859$ and $\lambda4339$ also contribute to the fluxes of \ha, \hb~and \hg, respectively.
The signal--to--noise ratio (S/N) of spectra obtained at $+1.40$ and $+1.42\,\rm{days}$ is not sufficient to rule out the presence of this feature also at earlier phases. 

\begin{figure}
\begin{center}
\includegraphics[width=\columnwidth]{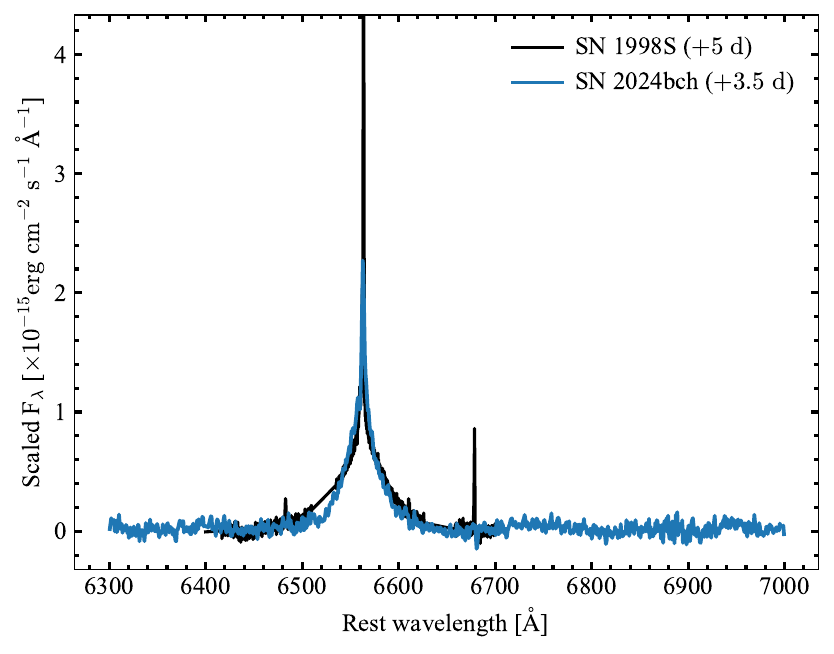}
\caption{Comparison of the higher-resolution spectrum available ($R\sim3500$) for \sn~obtained $\simeq+3.5\,\rm{days}$ with AFOSC with the one of SN~1998S (published by \citealt{2015ApJ...806..213S}) at a similar phase (adopting $\rm{JD}=2450872.5$ as reported by \citealt{2000MNRAS.318.1093F}). \label{fig:esLines}}
\end{center}
\end{figure}

\begin{figure}
\begin{center}
\includegraphics[width=\columnwidth]{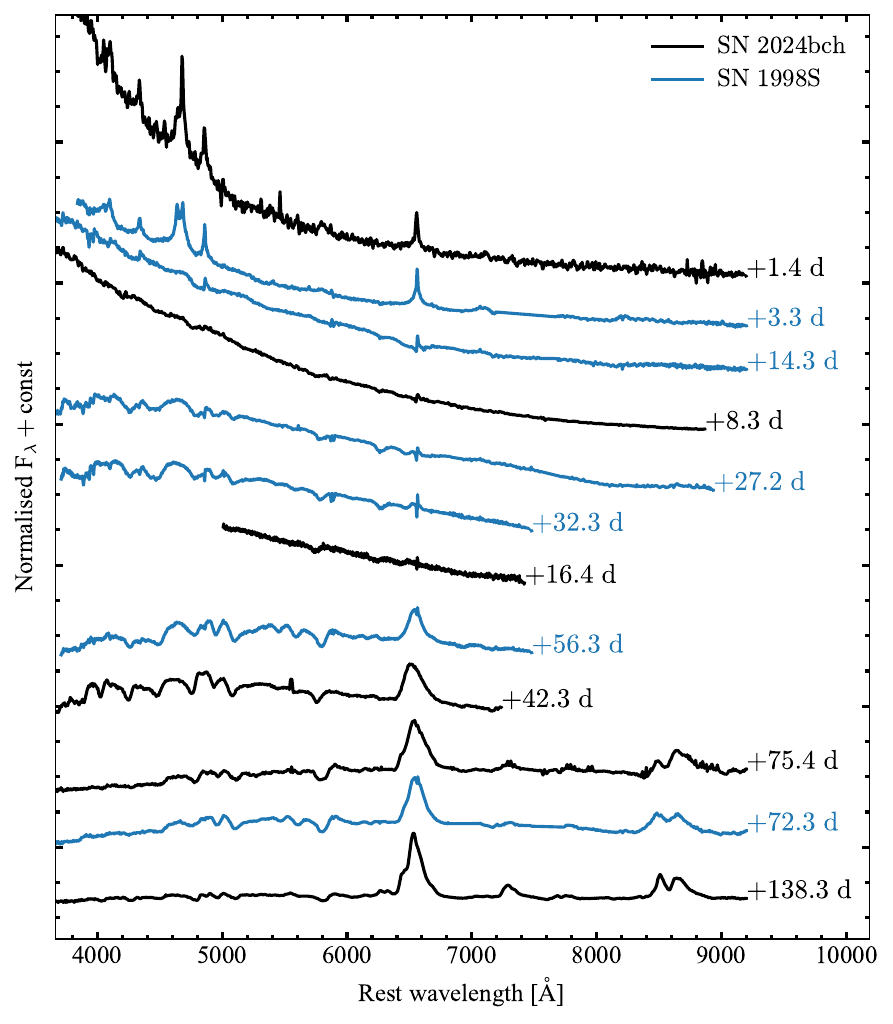}
\caption{Comparison between the spectroscopic evolution of \sn~and SN~1998S up to $\sim+140\,\rm{days}$ \citep[spectra from][]{2000ApJ...536..239L,2001MNRAS.325..907F}. \label{fig:specComp98S}}
\end{center}
\end{figure}

\begin{figure}
\begin{center}
\includegraphics[width=\columnwidth]{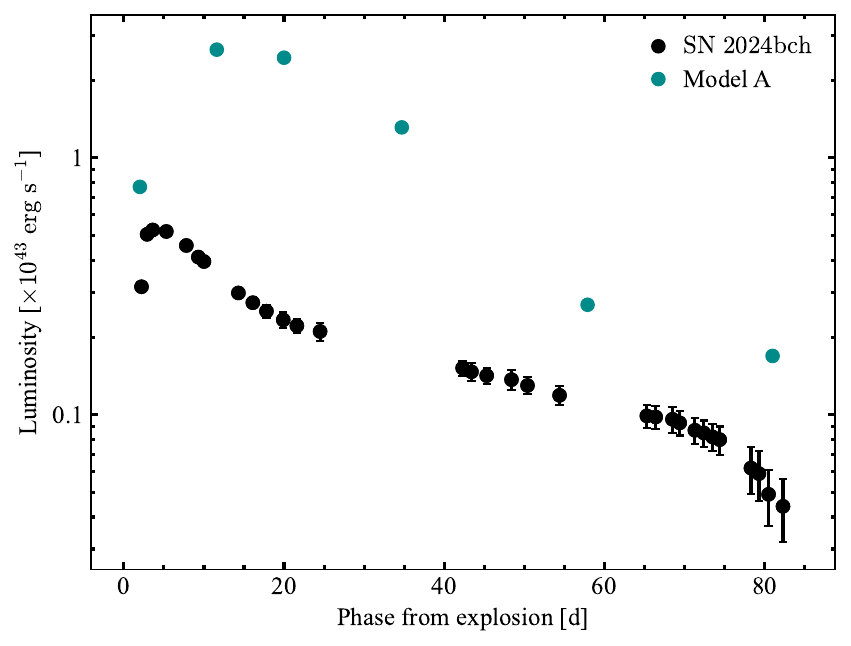}
\caption{Comparison between the bolometric light curve of \sn~and ``Model A" from \citet{2016MNRAS.458.2094D} built to explain the luminosity evolution of SN~1998S. The model fails to reproduce both the fast rise and the overall luminosity of \sn~suggesting ejecta-CSM interaction does not play a role in powering its light curve. \label{fig:bolComp98S}}
\end{center}
\end{figure}

In Fig.~\ref{fig:esLines}, we also compare the \ha~region in the medium resolution spectrum of \sn~($R\sim3500$) obtained at $+3.51\,\rm{days}$ with the one obtained at a similar epoch for SN~1998S \citep{2015ApJ...806..213S}.
\sn~shows a similar profile, with broad wings due to electron scattering in the unshocked circumstellar material, which also emits other narrow high-ionisation features visible up to $\simeq+8\,\rm{days}$.
Although broad wings in relatively narrow emission lines (as in the case of Type IIn SNe) have been often interpreted as an additional emission component from underlying fast-moving SN ejecta or shocked regions \citep[see, e.g.,][]{1993MNRAS.262..128T}, \citet{2001MNRAS.326.1448C} showed that multiple scattering of photons by free electrons are able to reproduce the whole line profile in SN~1998S without invoking additional components \citep[see also][for a more recent discussion on the process]{2018MNRAS.475.1261H}, adopting a Thomson optical depth of the shell of at least $\tau\sim3-4$ at this epoch (for an electron temperature $T_e\simeq(2.5-3.0)\times10^4\,\rm{K}$).
Based on the remarkable similarities between the two objects, we can therefore assume a similar optical depth for the circumstellar environment of \sn.
To further investigate the similarities between the two transients, we compare their spectroscopic evolution up to $\simeq140\,\rm{days}$ in Fig.~\ref{fig:specComp98S}. \sn~exhibits a more rapid evolution, with high-ionisation features disappearing within the first $\simeq8\,\rm{days}$.
The spectro-photometric evolution of SN~1998S was succesfully reproduced by \citet{2016MNRAS.458.2094D} through radiation hydrodynamics and radiative transfer modelling, based on a scenario where the explosion occurred within a dense medium and a significant fraction of the SN luminosity is powered by ejecta-CSM interaction.
In their ``model A”, typical SN ejecta ($\sim10$\msun, with an $E_k=10^{51}\,\rm{erg}$) collide with an outer dense shell formed through a wind expanding at $v_w=10^2$\kms, characterised by a mass loss rate of $0.1\,\rm{M_{\odot}}\,\rm{yr^{-1}}$.
This intense mass loss episode lasted $3.5\,\rm{yr}$ and ended $\sim3\,\rm{yr}$ before the SN explosion, producing the shell with an inner radius of $1.0\times10^{15}\,\rm{cm}$ extending out to $2.1\times10^{15}\,\rm{cm}$, where its density drops sharply, a configuration similar to that proposed by \citet{2004MNRAS.352.1213C} for the Type IIn SN~1994W.
Figures~\ref{fig:specComp98S} and \ref{fig:bolComp98S} show that, although this model is able to reproduce the spectroscopic evolution of both transients (with \sn~showing a faster spectroscopic evolution), it fails to match both the rise and the peak luminosity of \sn.
The faster rise time and fainter bolometric light curve of \sn~reinforce our claim that interaction has only a marginal, if any, contribution to its luminosity output.

Narrow high-ionisation lines in early spectra are often attributed to a confined shell of gas expelled during the very late evolutionary stages of the progenitor, rapidly overtaken by the expanding SN ejecta.
The engine providing the energetic photons responsible of the ionisation of this gas is typically identified in the conversion of ejecta kinetic energy into radiation, as happens in interacting transients \citep[see][and references therein]{2003LNP...598..171C}.
The early evolution of \sn, on the other hand, does not show clear signatures of ongoing ejecta-CSM interaction, with a rapidly evolving SED consistent with a freely expanding photosphere where the contribution of an extra source of energy is not required to model its early bolometric light curve (see Section~\ref{sec:photometry}).
In addition, while in most cases the strength of narrow features progressively declines, in \sn~we note a rapid increase in their integrated luminosity up to $\sim+2.44\,\rm{days}$ (except for \ha, where the maximum occurs at $+3.40\,\rm{days}$, suggesting a larger radius for the H-rich region).
After maximum, the luminosity of narrow features rapidly decline until $t\simeq+7\,\rm{days}$, when they fade below the continuum level.
As discussed in Sect.~\ref{sec:photometry}, the increase in the flux of narrow lines does not correspond to a similar evolution of the pseudo-continuum temperature, although optical spectra do show an apparent increase in their slope at $\lambda<5000$\ang. 
The same effect was observed in the early spectra of SN~1998S \citep[see][]{2000ApJ...536..239L} and explained by invoking a nonstandard extinction law or the presence of dust ``echoing" the SN light from earlier epochs.
As for SN~1998S, a single blackbody cannot reproduce the spectral continuum at $+2.4\,\rm{days}$, which would suggest a similar interpretation for \sn. 
While a nonstandard extinction law seems implausible due to the negligible reddening along the line of sight of \sn, a light echo may be supported by the peculiar evolution of the narrow lines, as discussed below.
We note, on the other hand, that the SED of \sn~is well reproduced by a blackbody if we include fluxes obtained from UV photometry, ruling out the need of a hotter source of energy or a nonstandard extinction law.
This also highlights the importance of UV data when the peak of the emission is at bluer/much bluer wavelengths than those covered by optical spectra.

We note all narrow lines are blueshifted, with peaks progressively shifting towards their rest wavelengths.
In \ha~(the most prominent line) the shift decreases from $\simeq240$ to $\simeq30$\kms~up to $+3.4\,\rm{days}$ and similar values are measured from the other emission lines, although these may be affected by the lower S/N and the contamination of other emission features, as discussed above.
This is not the expected evolution for an ionised CSM accelerated by the SN radiation.
Radiative acceleration has already been discussed for other objects \citep[e.g. SNe~2010jl, 2015da and 2023ixf][]{2014ApJ...797..118F,2020A&A...635A..39T,2023ApJ...956...46S,2024Natur.627..759Z} and it can be recognised by a progressive increase in the blueshift of the line peaks.
Since in \sn~we see the opposite, we can assume a negligible acceleration of the CSM by the SN radiation field. The radius at which radiative acceleration does not efficiently affect the CSM velocity ($\sim$10$^{15}$ cm for \sn) may provide a lower limit for the inner CSM radius.
In gas shells of such size, light travel time and light echoes become both relevant in shaping the observed spectral features.
The evolution of narrow peaks towards their rest wavelengths in \sn, may then be interpreted as the contribution of emitting regions with progressively smaller velocity components along the line of sight.
In a CSM geometry similar to that presented by \citet[][see, e.g., their Fig.~6]{Miller2006gy_lightecho}, at early phases one would detect narrow features emitted by the region closest to Earth, blueshifted by $\simeq240$\kms~due to its motion towards the observer. As time progresses, the observer would be reached by photons emitted by more distant regions, with a progressively smaller velocity component along the line of sight. Consequently, at later phases, the emission peaks would be blueshifted only by $\simeq30$\kms.
This scenario would also explain the luminosity evolution of narrow lines, due to the progressive increase of the emitting volume.
The luminosity peak would in fact be expected when the emission is dominated by light from the inner shell layers on the opposite side with respect to the observer.
This is because we expect the inner part of the shell to be the most affected by the SN light, hence the luminosity of narrow features is likely dominated by photons emitted in this region.
In this context, the epoch at which narrow features reach their maximum luminosity can also be used to infer an independent estimate of the inner CSM radius through the light travel time ($R_{in}\simeq c\times t^{}/2$).
Adopting $t_{max}=3.4\,\rm{days}$ (the maximum luminosity measured for \ha), we obtained $R_{in}\simeq4.4\times10^{15}\,\rm{cm}$, in agreement with the lower limit derived above.
Typical SN ejecta (i.e., with $v_{ej}=10^4$\kms) would then reach $R_{in}$ not earlier than $t_{reach}\simeq+51\,\rm{days}$, $+68\,\rm{days}$ assuming $v_{ej}\simeq7500$\kms~adopted for the bulk of the ejecta in our model (see Section~\ref{sec:modelling}).
These phases are in agreement with the later spectroscopic evolution of \sn, which at $t\simeq+69\,\rm{days}$ shows possible signatured ejecta-CSM interaction in terms of a blue boxy component in the \ha~profile (see the discussion below and in Section~\ref{sec:conclusions}).
This analysis points towards a distant CSM, reached by the SN ejecta only several weeks after explosion and highlights the need of a source of ionising photons different than interaction for the early narrow features observed in \sn.

The Bowen fluorescence mechanism \citep{1934PASP...46..186B,1935ApJ....81....1B} may explain the presence of high-ionisation lines without involving a significant contribution of ejecta-CSM interaction. This mechanism relies on a fortuitous coincidence in wavelengths allowing \ion{He}{II} $\lambda303.783$ resonance line to populate the \ion{O}{III} $\rm{2p3d}^3\rm{P}_2$ level, followed by the emission of \ion{O}{III} $\lambda303.799$ \citep[O1 process; see][]{2006agna.book.....O}. The excited \ion{O}{III} can then emit several Bowen lines, including prominent ones at $\lambda3444$, $\lambda3133$, $\lambda3341$, $\lambda3312$, $\lambda3047$ and $\lambda3760$, as well as other extreme ultraviolet lines \citep[see the Grotrian diagrams reported in][]{1980ApJ...242..615K,2007A&A...464..715S}. Notably, \ion{O}{III} $\lambda374.436$ can further excite nitrogen through a resonance doublet at $\lambda374.434$ and $\lambda374.441$, populating the \ion{N}{III} $3\rm{d}^2\rm{D}_{3^{}/2,\,5^{}/2}$ levels, subsequently decaying to $3\rm{p}^{2}\rm{P^O}_{3/2}$ and $3\rm{p}^{2}\rm{D^O}_{1^{}/2}$ emitting \ion{N}{III} $\lambda4641$, $\lambda4641$ and $\lambda4642$, respectively, and to the $3\rm{s}^2\rm{S}_{1^{}/2}$ level emitting \ion{N}{III} $\lambda4097$ and $\lambda4103$ \citep[see Fig. 13 and the discussion in][]{2007A&A...464..715S}.
This mechanism was invoked to explain \ion{N}{III} lines in a variety of gaseous nebulae, including planetary nebulae, X-ray binaries, symbiotic stars and novae in the early nebular stages \citep[see][]{1996MNRAS.279.1137K} and, more recently, in the tidal disruption events (TDEs) iPTF15af \citep{2019ApJ...873...92B} and iPTF16fnl \citep{2019MNRAS.489.1463O} and may be a viable excitation mechanism for \sn~as well. 
\citet{2002ApJ...572..350F} used near-UV spectra to quantify the contribution of the Bowen mechanism in the Type IIn SN~1995N through \ion{O}{III} $\lambda3047$, $\lambda3133$, $\lambda3340$ and $\lambda3444$ lines, as well as the $I$(\ion{O}{III} $\lambda3132$)/$I$(\ion{He}{II} $\lambda4686$) ratio to estimate the Bowen yield for \ion{He}{II} $\lambda304$.
Unfortunately, our early spectra do not extend below $\sim3600$\ang~(see Table~\ref{tab:speclog} and Fig.~\ref{fig:earlyIDSpec}) and we cannot quantify the Bowen yield using \ion{O}{III} $\lambda3132$.
While early spectra of \sn~lack \ion{O}{III} $\lambda3444$ \citep[a key diagnostic of the Bowen mechanism also in planetary nebulae; see, e.g.,][]{1969ApJ...157.1201W,1993MNRAS.261..465L}, the presence of \ion{N}{III} $\lambda4641$ and \ion{He}{II} $\lambda4686$ in our observations aligns with Galactic X-ray sources \citep[e.g.,][]{1975ApJ...198..641M}.
Although these lines can also arise via dielectronic recombination \citep{1973ApJ...179..827M}, theoretical \citep[e.g.,][]{1976ApJ...206..847H} and observational \citep[e.g.,][]{1978ApJ...222L..33M} evidence strongly supports Bowen fluorescence as the dominant mechanism for amplifying \ion{N}{III} emission.
Thus, despite the absence of direct UV diagnostics, the observed spectral features suggest that Bowen fluorescence remains a viable explanation for \sn.

At $t>+7\,\rm{days}$ spectra are dominated by a blue ($T\simeq1.3\times10^4\,\rm{K}$), almost featureless continuum typically observed during the early phases of Type II SNe.
The persistence of a narrow \ha~line with an absorption component visible at $+7$ and up to $+16\,\rm{days}$, on the other hand, suggests the presence of a H-rich unshocked shell with a blue-velocity-at-zero-intensity (BVZI) $\simeq300$\kms~and a photospheric velocity $\simeq150$\kms~(as measured from the highest resolution spectrum available at these phases, obtained at $+16\,\rm{days}$); see the inset in Figure~\ref{fig:earlyIDSpec}.
Adopting the $R_{in}$ derived above and BVZI for the expansion velocity of the CSM, this would date the mass loss event producing the outer shell $\sim5\,\rm{years}$ before the SN explosion.

\begin{figure*}
\begin{center}
\includegraphics[width=\linewidth]{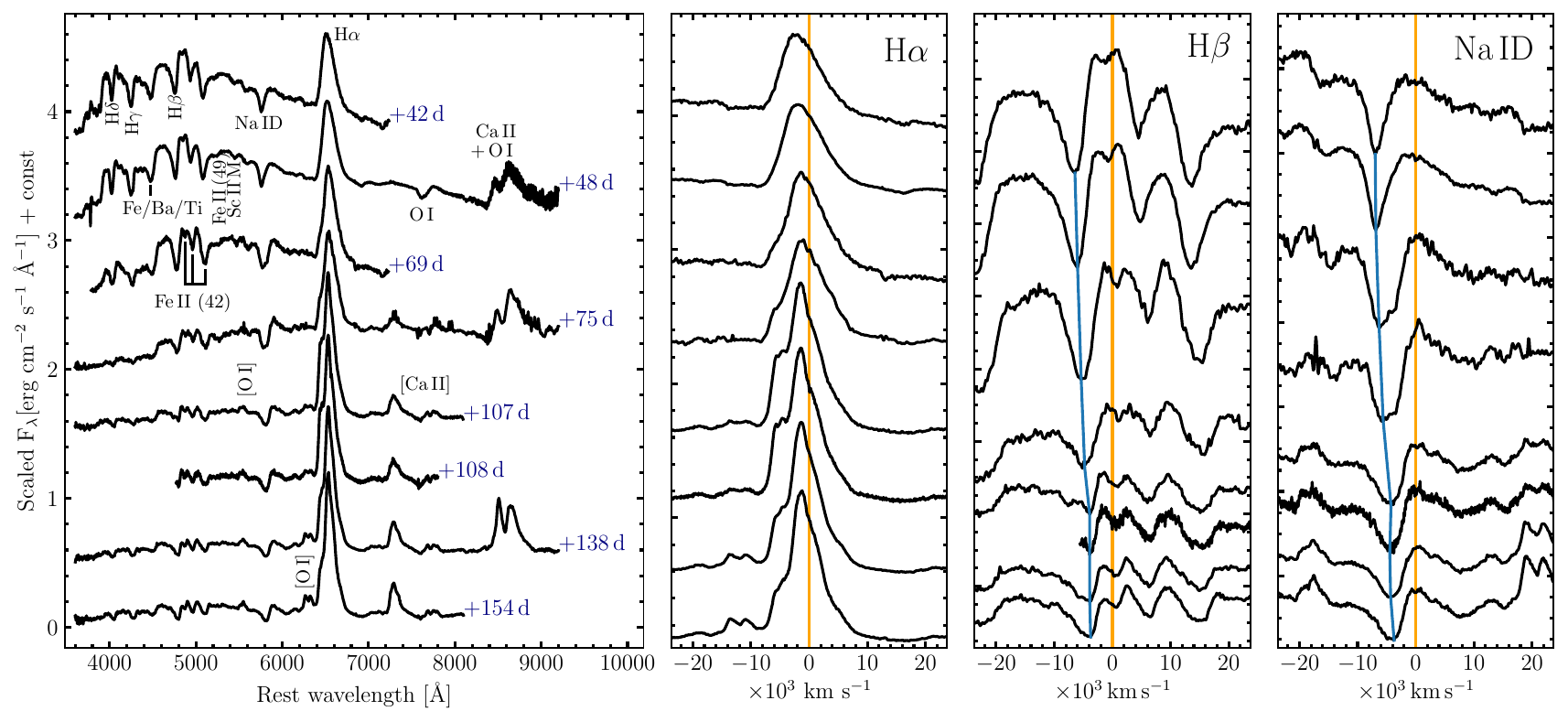}
\caption{From left to right: optical spectroscopic evolution of \sn~at $t\geq+42\,\rm{days}$ ({\bf Panel 1}; along with the identification of the most prominent spectral features), evolution of the \ha~profile ({\bf panel 2}), \hb~({\bf panel 3}) and \ion{Na}{ID} {\bf Panel 4} features in the velocity plane at the same epochs. Yellow lines correspond to the rest-frame wavelengths, while blue lines mark the evolution of the P Cygni absorption minima. Phases refer to the estimated explosion epoch. \label{fig:photoSpec}}
\end{center}
\end{figure*}

At $+42\,\rm{days}$ spectra show a significant metamorphosis, as broad P Cygni features begin to shape the spectral continuum. 
We identify broad absorption features corresponding to H Balmer lines (\hb, \hg~and \hd), \ion{He}{I}/\ion{Na}{ID} and \ion{O}{I} (from $+48\,\rm{days}$; see Fig.~\ref{fig:photoSpec}). 
The blue part of the continuum is also shaped by several features corresponding to the Fe-group elements, such as \ion{Fe}{II} (multiplet 42), \ion{Sc}{II} (multiplets 28 and 29) and \ion{Ba}{II} $\lambda6142$, where multiplets numbers are given following the notation reported in \citet{1945CoPri..20....1M}.
At $+48\,\rm{days}$, we also identify broad features corresponding to the \ion{Ca}{II} near infrared 8498, 8542, 8662\ang~triplet, \ion{O}{I} $\lambda8446$ and \ion{O}{I} $7772-7775$, although we cannot rule out their presence at earlier phases due to the limited spectral coverage of previous data. 
From the minima of \hb~and \ion{Na}{ID} at $+42\,\rm{days}$, we estimate photospheric expansion velocities for the outer ejecta layers of $6745\pm760$\kms~and $6830\pm740$\kms, respectively, with uncertainties computed as in \citet{2017ApJ...850...90G}.
Both values are higher than those typically observed in Type II SNe at similar phases \citep[see][]{2017ApJ...850...90G}.
At $+108\,\rm{days}$ the estimated expansion velocities are $3440\pm750$ and $4600\pm250$\kms~for \hb~and \ion{Na}{ID}, respectively, still above the average values inferred by \citet{2017ApJ...850...90G}.
During the same period, emission peaks (e.g., \ha, \hb) progressively shift towards the corresponding rest wavelengths (see panels 2 and 3 in Figure~\ref{fig:photoSpec}). 
As it is believed to form closer to the SN photosphere, another estimate of the photospheric expansion velocity can be inferred from the P Cygni profiles of the \ion{Fe}{II} 42 multiplet, in particular the relatively isolated $\lambda5169$ line \citep[see, e.g., the discussion in][]{2006A&A...447..691D}. 
This gives an expansion velocity of $5210\pm740$\kms~at $+42\,\rm{days}$, declining to $3900\pm300$\kms~at $+108\,\rm{days}$. 
These are both significantly above the mean velocities inferred by \citet{2017ApJ...850...90G} for their sample of Type II SNe at similar phases ($3760\pm1045$\kms~at $t\simeq+42\,\rm{days}$ and $2625\pm457$\kms~at $t\simeq+108\,\rm{days}$).

\subsection{Early nebular spectra} \label{sec:nebularSpec}
\begin{figure}
\begin{center}
\includegraphics[width=\columnwidth]{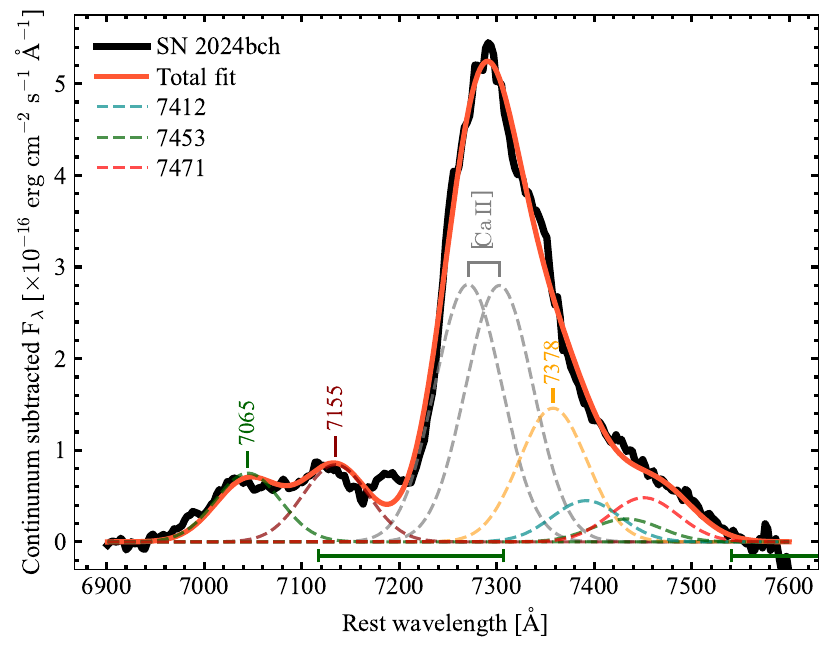}
\caption{Multi-Gaussian fit to the spectral region around \ion{[Ca}{II]} $\lambda\lambda7291$, 7324, following the prescriptions of \citet{2015MNRAS.448.2482J} and \citet{2016MNRAS.462..137T}. The fit was performed on the spectrum obtained at $+154\,\rm{days}$.} \label{fig:caiiregion}
\end{center}
\end{figure}

Forbidden \ion{[Ca}{II]} $\lambda\lambda7291$, 7324 emerge at $t\gtrsim75\,\rm{days}$, although we cannot rule out their presence at earlier phases due to the limited wavelength coverage of the $+69\,\rm{days}$ spectrum.
Several forbidden lines may contribute to the flux in the $7000-7600$\ang~spectral region, including \ion{[Fe}{II]} and stable \ion{[Ni}{II]}. 
At $+154\,\rm{days}$, \ion{[Ca}{II]} $\lambda\lambda7291$, 7323 shows a strongly asymmetric profile where \ion{[Fe}{II]} $\lambda7155$ likely contributes to the blue part of the emission, while \ion{[Ni}{II]} $\lambda7378$ and $\lambda7412$ and \ion{[Fe}{II]} $\lambda7453$ shape the red tail of the emission feature.
\citet{2015MNRAS.448.2482J} and \citet{2016MNRAS.462..137T} modelled the similar feature observed in SNe~2012ec and 2014G adopting a multi-Gaussian fit to the entire region, considering the contribution of \ion{[Ca}{II]} $\lambda\lambda7291$, 7323, \ion{[Fe}{II]} $\lambda7155$, $\lambda7172$, $\lambda7388$, $\lambda7453$ and \ion{[Ni}{II]} $\lambda7378$ and $\lambda7412$ to the overall profile.
Since we were unable to reproduce the overall spectral feature using the aforementioned lines, even letting the line ratios vary within reasonable ranges, we had to include two additional emissions around $\sim7050$ and $\sim7450$\ang.
Although we tentatively identified the blue excess as \ion{He}{I} $\lambda7065$, we could not find a reasonable identification for the red feature.
We therefore fixed the wavelength of the blue component and allowed the red one to vary within our fitting procedure.
In addition, to minimise the number of free parameters, we also ignored the contributions of \ion{[Fe}{II]} $\lambda7172$ and $\lambda7388$, which are relatively faint with respect to the other components in the reference model \citep[$L_{7172}^{}/L_{7155}=0.24$ and $L_{7388}^{}/L_{7172}=0.74$; see][]{2014MNRAS.439.3694J,2015MNRAS.448.2482J}.
Our ``best-fit” model is shown in Fig.~\ref{fig:caiiregion}, where we found the red extra component to be centered at $\lambda7471$ with a luminosity $(2.03\pm0.84)\times10^{38}$\ergs, while for \ion{He}{I} $\lambda7065$ we derived $(3.0\pm1.2)\times10^{38}$\ergs, corresponding to $L_{7471}^{}L_{7155}\simeq0.6$ and $L_{7065}^{}L_{7155}\simeq0.8$.
Following \citet{2014MNRAS.439.3694J}, we fixed the luminosity ratios of lines coming from the same levels, such as $L_{7453}^{}/L_{7155}=0.31$, and, as in \citet{2015MNRAS.448.2482J}, we also assumed $L_{7412}^{}/L_{7378}=0.31$.
Allowing the blueshift and FWHM of emission lines to vary (while forcing it to the same value for all lines), we inferred: $L_{7155}=(3.4\pm1.4)\times10^{38}$\ergs, $L_{7453}=(1.05\pm0.43)\times10^{38}$\ergs, $L_{7378}=(6.0\pm2.5)\times10^{38}$\ergs~and $L_{7412}=(1.87\pm0.77)\times10^{37}$\ergs, assuming $L_{7291}=L_{7323}=(1.15\pm0.47)\times10^{39}$\ergs, where we considered a distance of $19.9\pm4.1\,\rm{Mpc}$ and errors are dominated by its uncertainty ($\sim50$\% of the total uncertainty). 
The fitting procedure also provided a rigid shift $\Delta\lambda=20.10\pm0.52$\ang~($\simeq830$\kms~with respect to 7250\ang, the centre of the spectral region considered) and a FWHM velocity of $3332\pm45$\kms, both assumed to be constant for all lines \citep[as in][]{2016MNRAS.462..137T}.

Although the reference model was specifically designed for SN~2012aw \citep[see][for details]{2014MNRAS.439.3694J} and assumes local thermodynamic equilibrium (LTE) and that all lines are optically thin, it is capable to reproduce the observed spectral region between 6900 and 7600\ang~of \sn~as well, although we had to include two additional features (\ion{He}{I} $\lambda7065$ and $\lambda7471$). 
On the other hand, we note a red excess with respect to the same model was also observed in the \ion{[Ca}{II]} region of SNe~2012ec and 2014G \citep[][respectively]{2015MNRAS.448.2482J,2016MNRAS.462..137T}, so the red feature we find at $\lambda7471$ is likely real.
The luminosity ratio of the \ion{[Ni}{II]} $\lambda7378$ and \ion{[Fe}{II]} $\lambda7155$ lines can be used to estimate iron and nickel abundances, which, in LTE and assuming optically thin emission, can be expressed as follows:
\begin{equation}
\frac{L_{7378}}{L_{7155}}=4.9\,
\frac{n_{\mathrm{Ni\,II}}}{n_{\mathrm{Fe\,II}}}\,\exp\left(\frac{3250\,\rm{K}}{T}\right),
\label{eq:niferatio}
\end{equation}
where we used a ratio between partition functions $Z_{\rm{Ni\,II}}^{}/Z_{\rm{Fe\,II}}=0.25$, statistical weights $g^{\rm{Ni\,II}}_{\rm{4s2SF7/2}}=8$ and $g^{\rm{Fe\,II}}_{\rm{3d7a2G9/2}}=10$, transition probabilities $A_{7378}=0.23\,\rm{s^{-1}}$ and $A_{7155}=0.146\,\rm{s^{-1}}$, as in \citet{2015MNRAS.448.2482J}.
Although we lack evidence supporting LTE at these phases, this ratio is relatively insensitive to temperature and density, implying that deviations from LTE would similarly affect both lines.
However, we note that the fit of the \ion{[Ca}{II]} region accurately reproduces the overall shape of its spectral region, suggesting that large deviations from LTE are unlikely.
In addition, as shown by \citet{2017hsn..book..795J}, modelling of commonly observed SN lines suggests that the assumption of LTE may hold for \ion{[Ca}{II]} $\lambda\lambda7291$, 7324, \ion{[Fe}{II]} $\lambda7155$ and \ion{[Ni}{II]} $\lambda7378$ up to $\sim280-350\,\rm{days}$ after explosion.
Moreover, the assumption of optically thin emission may be reasonable at $+154\,\rm{days}$ (e.g., see the typical $t_{thin}$ reported there for \ion{[Ni}{II]}).
Assuming $\ion{Ni}{II}^{}/\ion{Fe}{II}\simeq\rm{Ni}^{}/\rm{Fe}$, Eq.~\ref{eq:niferatio} then gives the $\rm{Ni}^{}/\rm{Fe}$ production rate as a function of temperature.
If all $^{56}\rm{Ni}$ decayed and most of $^{56}\rm{Co}$ decayed into $^{56}\rm{Fe}$ ($\simeq65$\% at $+154\,\rm{days}$, assuming e-folding times of 8.8 and $111.4\,\rm{days}$ for $^{56}$Ni and $^{56}$Co, respectively), the Fe-zone temperature can be determined comparing the measured $L_{7155}^{}/M_{\rm{^{56}Ni}}$ to the theoretical value:
\begin{equation}
\frac{L_{7155}}{M_{\rm{^{56}Ni}}}=\frac{A_{7155}h\nu g^{\rm{Fe\,II}}_{\rm{3d7a2G9/2}}}{56m_uZ_{\rm{Fe\,II}}(T)}\,\exp\left(\frac{-22745\,\rm{K}}{T}\right),
\label{eq:Fe2Niratio}
\end{equation}
where we adopted a partition function $Z_{\rm{Fe\,II}}=15+0.006T$ as in \citet{2015MNRAS.448.2482J}.
The resulting temperature is $3946^{+85}_{-80}\,\rm{K}$, where, in comparing the two sides of Eq.~\ref{eq:Fe2Niratio}, we took into account the uncertainties on the derived $L_{7155}$ and $M_{\rm{^{56}Ni}}$ (see Sect.~\ref{sec:photometry}), but we did not consider possible contamination by primordial Fe and Ni \citep[see, e.g.,][]{2012MNRAS.420.3451M}, which may lead to underestimate the derived $\rm{\ion{Ni}{II}}^{}/\rm{\ion{Fe}{II}}$ ratio.
This temperature gives a $\rm{\ion{Ni}{II}}^{}/\rm{\ion{Fe}{II}}=0.16\pm0.06$, $\lesssim3$ times higher than the solar value of 0.056 \citep{2003ApJ...591.1220L}, in agreement with the values found for SNe~2012ec ($0.19\pm0.07$) and 2014G ($0.18\pm0.02$) by \citet{2015MNRAS.448.2482J} and \citet{2016MNRAS.462..137T}, respectively.
This is also higher than typical values found for other CC SNe, although the number of objects with estimated $\rm{\ion{Ni}{II}}^{}/\rm{\ion{Fe}{II}}$ is still relatively small \citep[see][]{1988Natur.331..505R,1989ApJ...342..364M,2007AJ....133...81M,1993ApJS...88..477W,2012A&A...546A..28J,2015MNRAS.448.2482J}. \\

\begin{figure}
\begin{center}
\includegraphics[width=\columnwidth]{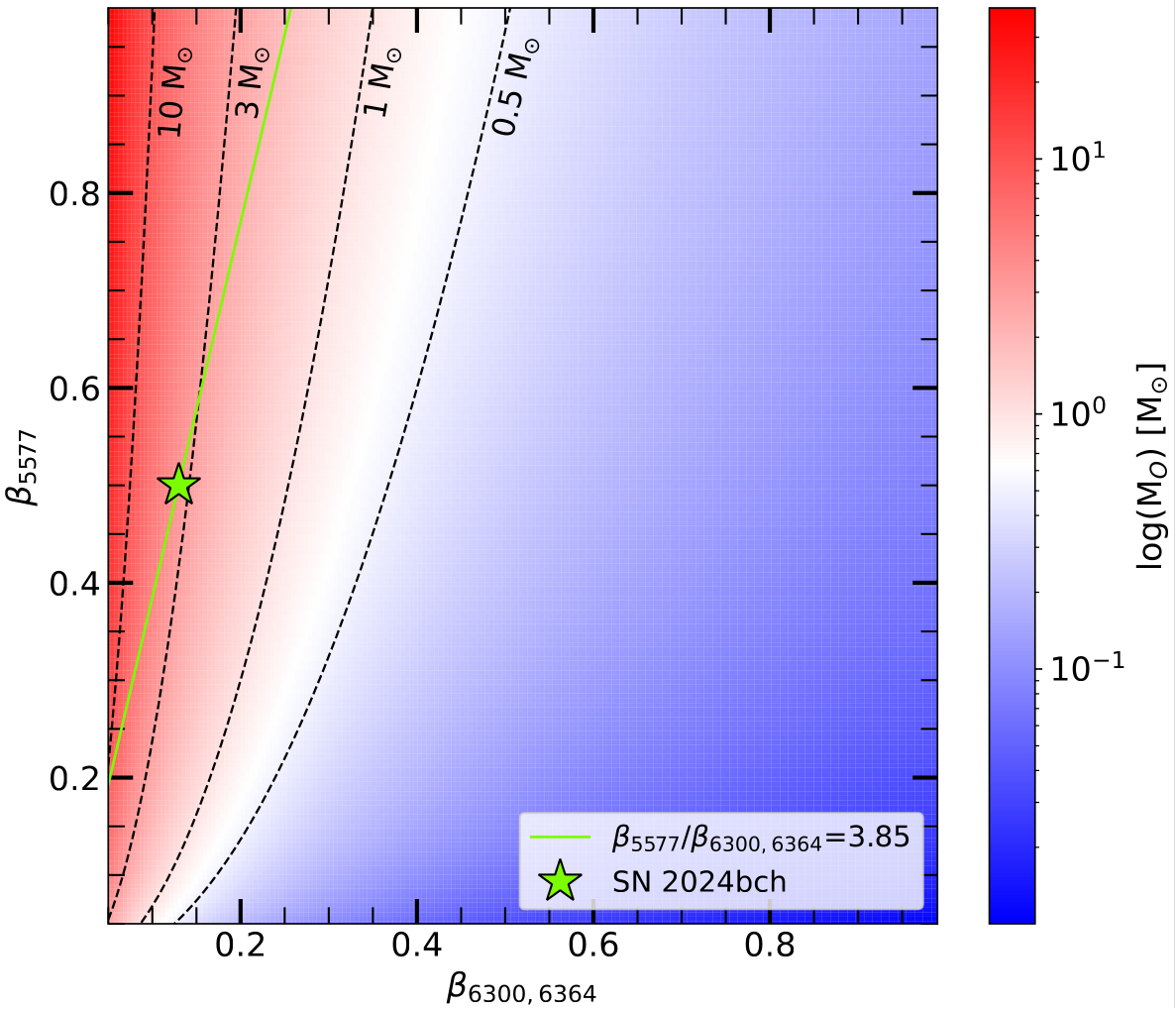}
\caption{Colour map showing the dependence of the O mass on the escape probabilities $\beta_{6300,6363}$ and $\beta_{5577}$, according to Eqs.~\ref{eq:oxTemp} and \ref{eq:oxmass}. Dashed black lines trace the location of some relevant O mass values. The green solid line displays the constant $\beta_{5577}/\beta_{6300,6363}$ ratio found through Eq.~\ref{eq:oxTemp}. The position of \sn~in this plane, based on the measurements performed on the spectrum at $+154\,\rm{days}$, is marked with a green star.} \label{fig:OmassBeta}
\end{center}
\end{figure}

At $t\geq107\,\rm{days}$, \ion{[O}{I]} $\lambda\lambda6300$, 6364 and 5577 emerge, progressively increasing their strength with respect to the local continuum.
At $+154\,\rm{days}$, the total luminosity of the $\lambda\lambda6300$, 6364 doublet is $\simeq1.4\times10^{39}$\ergs, while the luminosity ratio is $\sim1$, suggesting optically thick emission. 
While both lines are clearly visible, a fit to the overall profile is complicated by the complex region around the doublet, where the level of the continuum cannot be determined easily.
\ion{[O}{I]} $\lambda5577$ is usually weak in CC SNe, although we clearly detect the line in the early nebular phase of \sn.
In order to avoid contamination from nearby emission features, \citet{2014MNRAS.439.3694J} fit the overall profile including the contribution of \ion{[Fe}{II]} $\lambda5528$.
Following the same approach, we found a two-component Gaussian fit is able to reproduce the entire emission feature at $+154\,\rm{days}$, corresponding to \ion{[Fe}{II]} $\lambda5528$ and \ion{[O}{I]} $\lambda5577$ both with a blueshift comparable to the one measured from the multi-Gaussian fit to the $7000-7600$ spectral region ($20.10\pm0.52$\ang; see above).
The derived luminosity $L_{5577}\simeq3.0\times10^{38}$\ergs~can be used, along with the estimate of $L_{6300,6364}$ mentioned above, to derive the temperature of the emitting region.
In LTE:
\begin{equation}
\frac{L_{5577}}{L_{6300,6364}}=38\exp\left(\frac{-25790\,\rm{K}}{T}\right)\,\frac{\beta_{5577}}{\beta_{6300,6364}},
\label{eq:oxTemp}
\end{equation}
where $\beta_{5577}$ and $\beta_{6300,6364}$ are the photon escape probabilities in the Sobolev approximation \citep[$\beta_{\lambda}=(1-e^{-\tau_{\lambda}})^{}/\tau_{\lambda}$;][]{1957SvA.....1..678S}.
When the \ion{[O}{I]} doublet transitions to optically thin conditions, the relative intensities approach their intrinsic ratio of 3:1 since both transitions share the same upper energy level.
In approaching the optically thin regime (i.e. $\tau\lesssim 1$ and \ion{[O}{I]} $\lambda6300$ stronger than \ion{[O}{I]} $6364$), $\beta_{6300,6364}$ becomes $\gtrsim0.6$.
This is not the case for \sn~at $+154\,\rm{days}$, where the doublet is still in the optically thick regime.
On the other hand, assuming similar temperatures for the Fe-- and O--rich zones ($\simeq3946\,\rm{K}$, as derived above), we can use Eq.~\ref{eq:oxTemp} to estimate the escape probabilities ratio, resulting in $\beta_{5577}^{}/\beta_{6300,6364}\simeq3.85$. The dependence of the O mass on the photon escape probabilities is represented in Fig.~\ref{fig:OmassBeta}.

The oxygen yield of the explosion can be derived using:
\begin{equation}
M_{\rm{O\,I}}=\frac{L_{6300,6364}\,\beta^{-1}_{6300,6364}}{9.7\times10^{41}\,\rm{erg}\,\rm{s^{-1}}}\exp\left(\frac{22720\,\rm{K}}{T}\right)\,\rm{M_{\odot}},
\label{eq:oxmass}
\end{equation}
where we adopted the atomic constants reported in \citet{2014MNRAS.439.3694J}.
Substituting the measured $L_{6300,6364}$ in Eq.~\ref{eq:oxmass} gives $M_{\ion{O}{I}}=0.46\,\beta^{-1}_{6300,6364}$ assuming a similar temperature for the Fe-- and O--rich zones. 
As an example, the photospheric temperature at $+158\,\rm{days}$ inferred in Sect.~\ref{sec:photometry} (i.e., considering the O--rich zone to be closer to the photosphere than the Fe--rich zone), would give $M_{\ion{O}{I}}=0.27\,\beta^{-1}_{6300,6364}$.
Assuming $\beta_{6300,6364}=0.32$ (equivalent to $\tau_{6300,6364}\simeq3$) would give $M_{\ion{O}{I}}\simeq0.8-1.4$\msun, consistent with a $M_{ZAMS}=15-20$\msun~according to \citet{1997NuPhA.616...79N} and \citet{2002ApJ...576..323R}.
However, these estimates are heavily affected by the assumptions made, including the temperature of the emitting region, the value of $\beta_{6300,6364}$ and the optical thickness of the \ion{[O}{I]} $\lambda\lambda6300$, 6364 doublet at $+154\,\rm{days}$.
One of the most important sources of uncertainty is the assumed temperature of the O-rich region, reflected in Equation~\ref{eq:oxmass} which, in our case, is not supported by observational nor theoretical considerations.
As seen in Sect.~\ref{sec:nebularSpec}, an uncertainty of only $\sim400\,\rm{K}$ (the difference between $T_{Fe}$ and $T_{ph}$ at $+158\,\rm{days}$), translates to roughly a factor of two uncertainty in the derived oxygen mass, spanning from $0.27\beta^{-1}_{6300,6364}$ to $0.46\beta^{-1}_{6300,6364}$.
Later spectra (e.g., when the transition to the optically thin regime will occur; see Valerin et al. in preparation) will provide more precise constraints and allow comparison with these preliminary estimates.

\section{Summary and conclusions} \label{sec:conclusions}
In the previous sections we have detailed our analysis on the photometric (Sect.~\ref{sec:photometry}) and spectroscopic (Sect.~\ref{sec:spectroscopy}) evolution of the Type II \sn. 

\begin{figure}
\begin{center}
\includegraphics[width=\columnwidth]{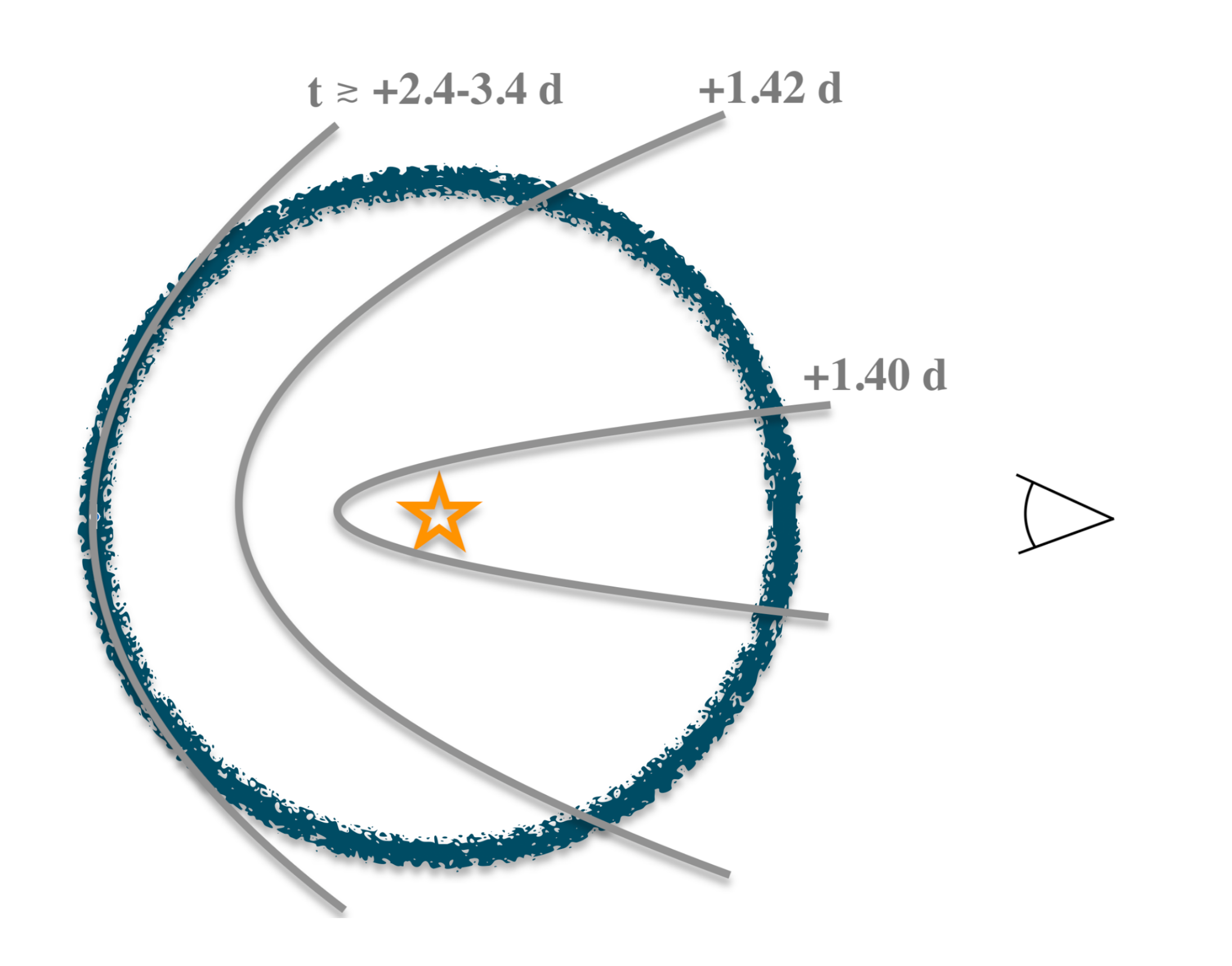}
\caption{Schematic view, not to scale, of the preferred geometrical configuration for the CSM of \sn. An outer shell with $R_{in}\gtrsim4.4\,\rm{10^{15}\,\rm{cm}}$ is reached by radiation at different times and emits high-ionisation features through the Bowen fluorescence mechanism. This scenario can simultaneously explain the evolution of the luminosities and redshifts of narrow features without the need of ejecta-CSM interaction as additional source of energy. \label{fig:CSMcartoon}}
\end{center}
\end{figure}

Early spectra ($t\lesssim+7\,\rm{days}$) are dominated by narrow emission lines such as \ion{He}{II}, \ion{N}{III}, \ion{C}{IV} in addition to those of the H Balmer series (from \ha~to \hd).
These show broad wings due to Thomson scattering by free electrons in a dense CSM \citep[see, e.g.,][and the comparison with the \ha~profile of SN~1998S at sufficiently high resolution in Fig.~\ref{fig:esLines}]{2001MNRAS.326.1448C,2018MNRAS.475.1261H} with an integrated luminosity increasing up to a maximum occurring $\simeq2-3\,\rm{days}$ after explosion for all narrow lines. 
As narrow lines increase in strength, their peaks progressively shift towards their rest wavelengths. 
We interpreted this evolution as a geometrical effect of the emitting region, in a similar fashion to a light echo \citep[see the geometry described for SN~2006gy in Fig. 6 by][and Figure~\ref{fig:CSMcartoon}]{Miller2006gy_lightecho}.
At the time of the \ha~maximum luminosity ($\simeq+3.4\,\rm{days}$), we inferred the inner radius of the H-rich circumstellar shell only considering the light travel time, resulting in an inner radius $R_{in}\approx4.4\times10^{15}\,\rm{cm}$.
This can be reached by typical SN ejecta (with $v_{ej}\simeq10^4$\kms; consistent with the BVZI measured from the \hb~P Cygni profile) not earlier than $t_{reach}\simeq51\,\rm{days}$, corresponding to the appearance of a blue shoulder in the \ha~line profile, occurring at $t\gtrsim+69\,\rm{days}$.
It is therefore unlikely that the disappearance of narrow high-ionisation emission lines at $\simeq+7\,\rm{days}$ (see Fig~\ref{fig:earlyIDSpec}) is due to the CSM being swept away by the SN ejecta.
In addition, the contribution of CSM-ejecta interaction to the luminosity output of the transient could be marginal and it is not necessary to reproduce the observed data of \sn, as shown by the modelling of its bolometric light curve (see Sect.~\ref{sec:photometry} and Figure~\ref{fig:ModelBolom}). 
Our ``best-fit" model is able to reproduce the entire evolution of the bolometric luminosity invoking the contribution of two regions: a bulk of relatively massive ejecta ($M_{bulk}=5$\msun), whose emission accounts for the plateau luminosity, and an extended envelope ($M_{env}=0.2$\msun) with an outer radius $R_{env}=7\times10^{13}\,\rm{cm}$, responsible of the early luminosity peak due to its short diffusion time.
This challenges the most common scenario invoked for SNe displaying ``flash spectroscopy features" \citep[][]{2014Natur.509..471G}, where shocks are the main source of ionising photons behind high-ionisation features.

Narrow high-ionisation features may instead be the result of fluorescence due to efficient absorption of extreme early UV radiation and subsequent resonance due to the nearly coincident wavelengths of \ion{He}{II}, \ion{O}{III} and \ion{N}{III} transitions.
H Balmer (and possibly \ion{He}{II} $\lambda4686$ and \ion{C}{IV}) lines may instead be interpreted as recombination features following the photo-ionisation of the CSM by the SN shock breakout, possibly extended by the presence of an inflated envelope \citep[see, e.g.,][]{2015A&A...575L..10M}.
All these features are typically observed in a significant fraction of CC SNe with sufficiently early spectra \citep[see][]{2021ApJ...912...46B} and this alternative scenario may be relevant in interpreting their early evolution.
In this context, narrow emission lines in SN spectra only prove the presence of an outer CSM and are not related to ``efficient" interaction \citep[see, e.g.,][for a definition of ``strongly interacting" SNe]{2023A&A...673A.127S}.
This has important implications also for multi-messenger astronomy, as strongly interacting SNe may provide a viable source of neutrinos in an energy range comparable to those detected by IceCube \citep[up to $1\,\rm{Pev}$; see][]{2020ApJ...904....4F,2023MNRAS.524.3366P}.
A lack of ``early interaction" may also explain non-detection of some SNe II with high-ionisation features at radio wavelengths at early times \citep[see, e.g.,][]{2021ApJ...907...52T}.

Possible signatures of CSM interaction appear at $t\gtrsim+69\,\rm{days}$, in terms of a blue ``shoulder" visible in the \ha~emission profile (see Figure~\ref{fig:photoSpec}).
However, according to the photometric analysis, this neither contributes significantly to the total energy output of the transient, nor causes a significant change in the ionisation state of the emitting region.
Alternatively, a flat-topped boxy profile may be produced by a recombining H-rich inner shell visible at later times after the SN photosphere recedes \citep[see, e.g.,][for an analytical discussion on possible line profiles emitted by expanding spherical shells]{1983A&A...120....6W,1987A&A...183..319B}.
This shell should contain a relatively small mass of H in order to account for the shape of the light curve, which at $+85\,\rm{days}$ already settles on the radioactive decay tail.
In any case, interaction of fast SN ejecta with the pre-existing CSM, if present at all, does not seem to contribute significantly to the total energy output of the transient at any phase. 
While the negligible contribution of interaction at early phases can be explained invoking Bowen fluorescence powering narrow lines at $t\lesssim+7\,\rm{days}$, explaining the lack of a significant extra source of energy at later phases is non-trivial and may involve a complex geometry or CSM structure.
The presence of a ``clumpy" CSM was discussed by \citet{1994MNRAS.268..173C} to explain the multi-component emission lines observed in SN~1988Z and similar and/or asymmetric structures seem to be common in eruptive/explosive transients \citep[see, e.g.,][]{2001AJ....121.1111S,2021A&A...655A..32K}, while a peculiar geometry was invoked to explain ``hidden interaction" in PTF11iqb, iPTF14hls and SN~2020faa \citep[see, e.g.,][]{2015MNRAS.449.1876S,2023A&A...673A.127S}.
The dataset collected so far does not allow us to further constrain the physical and geometrical properties of the SN explosion and its circumstellar environment.

Modelling of the radioactive tail (i.e. at $t\gtrsim+90\,\rm{days}$; see Fig.~\ref{fig:lateBoloFit}), reveals a non-complete trapping of the $\gamma-$rays produced in the $^{56}\rm{Ni}\rightarrow^{56}\rm{Fe}$ decay and results in a mass of radioactive Ni produced in the explosion of $\simeq0.048$\msun. 
Including information provided by early nebular emission features and assuming $\tau\simeq3$ for the O--rich region, this results in a mass of synthesised O of $0.8-1.4$\msun, which, according to the nucleosynthesis yields computed by \citet{1997NuPhA.616...79N} and \citet{2002ApJ...576..323R}, corresponds to a $M_{ZAMS}=15-20$\msun~progenitor.
This estimate is strongly limited by the estimated temperature of the O--rich region and the optical thickness of the \ion{[O}{I]} $\lambda\lambda6300$, 6364 doublet at $+154\,\rm{days}$, the last available spectroscopic epoch before the end of the transient visibility window and will be refined with proper spectroscopic data once the transient will be visible again.

Modelling of the bolometric light curve along with our analysis of spectroscopic data suggest that \sn~exploded expelling a moderate amount of H-rich gas ($\lesssim6$\msun) within an extended ($R_{in}\simeq4.4\times10^{15}\,\rm{cm}$) and likely massive shell of CSM. 
The presence of this outer CSM, along with the estimated $M_{ZAMS}$ for the progenitor and ejected mass from our modelling, support the claim that fast-evolving Type II SNe are produced by massive stars that lose a significant fraction of their outer H-rich layers before explosion. \\

\section*{Data Availability}
Table~\ref{table:bolom} and data shown in Figures~\ref{fig:lightCurves}, \ref{fig:earlyIDSpec}, \ref{fig:esLines} and \ref{fig:photoSpec} are only available in electronic form at the CDS via anonymous ftp to \url{cdsarc.u-strasbg.fr} (130.79.128.5) or via \url{http://cdsweb.u-strasbg.fr/cgi-bin/qcat?J/A+A/}.

\begin{acknowledgements}
We thank Elena Mason for the useful and lively discussions. \\
G.~V., A.~P., I.~S., S.~B. L.~Tom. and P.~O. acknowledge support from the PRIN-INAF 2022 project ``Shedding light on the nature of gap transients: from the observations to the model". F.~O. acknowledges support from MIUR, PRIN 2020 (grant 2020KB33TP) ``Multimessenger astronomy in the Einstein Telescope Era (METE)'' and from INAF-MINIGRANT (2023): ``SeaTiDE - Searching for Tidal Disruption Events with ZTF: the Tidal Disruption Event population in the era of wide field surveys". \\
A.~R. also acknowledges financial support from the GRAWITA Large Program Grant (PI P. D’Avanzo). \\
This research made use of the Spanish Virtual Observatory\footnote{\url{https://svo.cab.inta-csic.es}} project, funded by MCIN/AEI/10.13039/501100011033/ through grant PID2020-112949GB-I00, the NASA/IPAC Extragalactic Database (NED), funded by the National Aeronautics and Space Administration and operated by the California Institute of Technology and the HyperLeda database (\url{http://leda.univ-lyon1.fr}). \\
The ZTF forced-photometry service was funded under the Heising-Simons Foundation grant \#12540303 (PI: Graham). \\
Based on observations collected with the Copernico and Schmidt telescopes (Asiago, Italy) of the INAF -- Osservatorio Astronomico di Padova and the Wide-field Optical Telescope (WOT), a $67/91\,\rm{cm}$ Schmidt telescope equipped with an Apogee Aspen CG16M camera located at the Campo Imperatore observatory in l'Aquila (Italy) of the INAF -- Osservatorio Astronomico d'Abruzzo. \\
This article is also based on observations made in the Observatorios de Canarias del IAC with the Telescopio Nazionale Galileo, operated on the island of La Palma by INAF at the Observatorio del Roque de los Muchachos under the A47TAC\_37 (PI: G.~Valerin) and A49TAC\_64 (PI: L.~Tartaglia) programmes.

\end{acknowledgements}

\bibliographystyle{aa}
\bibliography{mybib}

\end{document}